\documentclass[a4paper,11pt]{article}
\usepackage{jcappub} 
\usepackage[utf8]{inputenc}
\usepackage{amsmath,amssymb,amsfonts}
\usepackage{geometry}
\usepackage{float}
\usepackage{graphicx}
\usepackage{subcaption}
\usepackage{caption}
\usepackage{booktabs} 
\usepackage{multirow} 

\usepackage[colorlinks=true,
            urlcolor=blue,
            anchorcolor=blue,
            citecolor=blue,
            filecolor=blue,
            linkcolor=blue,
            menucolor=blue,
            linktocpage=true]{hyperref} 

\title{\boldmath Late-Time Background Expansion and Cosmographic Signatures of Finsler-Randers-Sasaki Cosmology}

\author[a]{Thunga S.}
\author[b]{Devaraja Mallesha Naik}
\author[a]{Manaswini B. S.}
\author[a,1]{V. Venkatesha\note{Corresponding author.}}

\affiliation[a]{Department of P.G. Studies and Research in Mathematics,\\
Kuvempu University, Shankaraghatta, Shivamogga 577451, Karnataka, India}

\affiliation[b]{Department of Mathematics,\\
Government First Grade College Shikaripura, Shivamogga 577427, Karnataka, India}

\emailAdd{vensmath@gmail.com}

\abstract{
In this work, we investigate the late-time expansion history and geometric diagnostics of logarithmic and exponential Finsler--Randers--Sasaki (FRS) cosmological models. We constrain both scenarios using cosmic chronometers, Baryon acoustic oscillations, and Pantheon+ Type~Ia supernovae, considering baseline fits as well as configurations calibrated with the local SH0ES distance-ladder prior. In the uncalibrated case, both profiles provide consistent descriptions of the late-time expansion history. Incorporating the SH0ES prior increases the inferred Hubble constant and modifies the inferred acoustic scale, reflecting the interplay between the local distance-ladder calibration and the BAO distance scale. Cosmographic reconstructions reveal a smooth transition from deceleration to acceleration at intermediate redshifts. Over the observationally relevant redshift range, both models exhibit a non-phantom, quintessence-like effective dark-energy sector. Finally, statefinder diagnostics distinguish the two FRS geometries: the logarithmic model remains relatively close to the canonical flat $\Lambda\mathrm{CDM}$ fixed point, whereas the exponential model follows a distinct dynamical trajectory. These results demonstrate that, despite their similar background expansion histories, the two FRS constructions can exhibit substantially different higher-order kinematic signatures. }

\begin{document}
\maketitle
\flushbottom

\section{Introduction}
\label{sec:intro}

Standard general relativity identifies gravitation with the curvature of a four-dimensional pseudo-Riemannian manifold~\cite{Einstein1916,Will2014}. 
While this geometric formulation has withstood stringent Solar System experiments and direct gravitational-wave observations~\cite{Will2014,Abbott2016}, it confronts severe limitations at both ultraviolet and infrared limits. The non-renormalizable Einstein--Hilbert action operates consistently only as an effective field theory restricted to energy regimes below the Planck scale~\cite{Donoghue1994}. 
On cosmological horizons, the concordance $\Lambda\mathrm{CDM}$ model depends on a cold dark matter component and a cosmological constant $\Lambda$, yet their underlying microscopic mechanisms remain unidentified~\cite{Peebles2003,Copeland2006}. 
This theoretical ambiguity is sharpened by persistent observational discordances between early and late-universe probes, most conspicuously the tensions surrounding the local Hubble constant ($H_0$) and the matter clustering amplitude ($S_8$)~\cite{DiValentino2021a,DiValentino2021b,Abdalla2022,Perivolaropoulos2022}. 
Such empirical and foundational hurdles motivate extensions of gravitational theory.

The standard route to modifying gravity involves introducing additional dynamical fields, higher-order curvature invariants or non-Riemannian connections~\cite{Capozziello2011,CANTATA2021,Joyce2015}. Familiar implementations include scalar-tensor theories~\cite{Brans1961} and $f(R)$ gravity, where non-linear terms in the Ricci scalar drive early inflation or late-time cosmic acceleration~\cite{Starobinsky1980,Capozziello2002,Carroll2004}, culminating in Horndeski gravity and its derivative-coupled extensions that maintain second-order field equations to evade Ostrogradsky instabilities~\cite{Horndeski1974,Kobayashi2019}. Alternatively, gravity can be formulated without spacetime curvature within the ``geometric trinity,'' assigning gravitational interactions to torsion in teleparallel $f(T)$ gravity~\cite{Cai2016,Krssak2019} or non-metricity in symmetric teleparallel $f(Q)$ models~\cite{BeltranJimenez2018,BeltranJimenez2019,Heisenberg2024}. However, standard extensions built on classical Riemannian manifolds often face acute fine-tuning issues and severe degeneracies across astrophysical and cosmological regimes. Instead of appending external fields or ad-hoc couplings to an already constrained metric tensor, a more fundamental path is to broaden the geometry itself, turning naturally to metric-affine gravity~\cite{Hehl1995} and tangent-bundle geometries that incorporate directional degrees of freedom from first principles~\cite{Pfeifer2012,Hohmann2020,Kapsabelis2024}.

Finsler geometry generalizes the pseudo-Riemannian metric by defining it directly on the tangent bundle $TM$, introducing an explicit dependence on directional fiber coordinates $y^\alpha$ alongside position $x^\mu$~\cite{Pfeifer2012,Stavrinos2013}. To preserve coordinate covariance under this velocity dependence, one introduces a non-linear connection $N^\alpha_\mu$. This connection decomposes the tangent bundle into horizontal and vertical subbundles, $TTM = HTM \oplus VTM$, which naturally gives rise to generalized curvature and torsion forms~\cite{Bucataru2007,Vacaru2008}. 
In the Finsler--Randers--Sasaki (FRS) framework, this geometry is constructed by coupling a Randers metric perturbation~\cite{Randers1941} with a Sasaki lift onto $TM$~\cite{Sasaki1958,Miron2001}, defining an eight-dimensional Lorentz tangent bundle~\cite{Kapsabelis2024}. Once inserted into the modified Einstein-like field equations, the geometric contributions from the horizontal and vertical sectors act as an effective energy-momentum tensor on the spacetime base. Consequently, FRS cosmology accommodates a dynamical dark-energy behavior without requiring an explicit scalar field or cosmological constant, establishing background solutions that transition smoothly from matter domination to accelerated expansion~\cite{Kapsabelis2024}.

Observational validation of these geometric extensions requires confrontation with complementary datasets tracking cosmic expansion across distinct epochs. 
While baseline expansion dynamics are primarily anchored by the Hubble parameter $H(z)$, its local value $H_0$~\cite{Jimenez2002}, and the fractional matter density $\Omega_{m0}$~\cite{Planck2020}, higher-order kinematic derivatives yield model-independent diagnostic power. 
In particular, cosmographic coefficients specifically the deceleration ($q$), jerk ($j$), snap ($s$), and lerk ($l$) parameters reconstruct cosmic kinematics purely from successive time derivatives of the scale factor, without imposing an a priori dynamical model~\cite{Visser2004,Capozziello2019}. 
Complementing these kinematic series, the statefinder diagnostic pair $\{r, s\}$ and its evolutionary phase-space trajectories systematically discriminate modified gravitational frameworks from the concordance $\Lambda\mathrm{CDM}$ scenario~\cite{Sahni2003,Alam2003}. 
Furthermore, reconstructing the effective dark-energy density $\rho_{\mathrm{DE}}(z)$ alongside the dynamical equation-of-state parameter $w_{\mathrm{DE}}(z)$ directly maps the geometric modifications of the field equations onto observable cosmological phenomenology~\cite{Huterer2001,Sahni2006}.

In this paper, we explore how an osculating Finsler--Randers--Sasaki (FRS) metric over an eight-dimensional Lorentz tangent bundle modifies standard cosmological dynamics~\cite{Kapsabelis2024}. Projecting the horizontal field equations onto a flat Friedmann--Lema\^{\i}tre--Robertson--Walker spacetime yields revised Friedmann relations. Employing MCMC sampling across cosmic chronometers, Pantheon+, DESI DR2 BAO and SH0ES data, we place observational bounds on $H_0$ , $\Omega_{m0}$ and model parameters. Kinematic and geometrical tests spanning cosmography $(q, j, s, l)$, statefinder trajectories and the dark-energy equation of state are carried out alongside information-theoretic model comparisons ($\chi^2_{\mathrm{min}}$, AIC~\cite{Akaike1974} and BIC~\cite{Schwarz1978}) against $\Lambda\mathrm{CDM}$. 
 This article is organized as follows: Sec.~\ref{sec:theory} sets out the FRS framework, background expansion equations and cosmological diagnostics; Sec.~\ref{sec:cosmological_parametrizations} introduces the logarithmic and exponential cosmological parametrizations; Sec.~\ref{sec:data} details the observational datasets including cosmic chronometers, DESI DR2 BAO and Type Ia supernovae. Sec.~\ref{sec:results_discussion} reports our parameter constraints, cosmographic kinematic evolution, statefinder trajectories and statistical assessments and Sec.~\ref{sec:conclusions} provides concluding remarks.


\section{Finsler--Randers--Sasaki Cosmology and Cosmological Diagnostics}
\label{sec:theory}

Here, we present the geometrical foundations of the Finsler--Randers--Sasaki (FRS) space over an eight-dimensional Lorentz tangent bundle $TM$, adopting the construction developed by Kapsabelis, Saridakis and Stavrinos~\cite{Kapsabelis2024}. Building on this geometry, we subsequently establish the symmetry and background constraints necessary to track cosmic evolution in our setup.


\subsection{FRS Geometry and Background Expansion}
\label{subsec:frs_geometry_background}

We formulate the spacetime geometry over an eight-dimensional tangent bundle $T\mathcal{M}$ constructed upon a four-dimensional Lorentzian base manifold $(\mathcal{M}, g_{\mu\nu})$ with signature $(-,+,+,+)$. Introducing local coordinates $x^\mu$ ($\mu=0,1,2,3$) on the base alongside directional fiber velocities $y^\alpha \in T_x\mathcal{M}$, the bundle is mapped via the coordinate system $u^A = (x^\mu, y^\alpha)$~\citep{Vacaru2008,Pfeifer2012}. Within an osculating Finsler--Randers setting~\citep{Vacaru2012,Kapsabelis2024}, the fundamental metric function incorporates a weak anisotropic perturbation covector $A_\alpha(x)$ ($\|A_\alpha\| \ll 1$):
\begin{equation}
F(x, y) = \sqrt{-g_{\mu\nu}(x)y^\mu y^\nu} + A_\alpha(x)y^\alpha.
\label{eq:finsler_metric_function}
\end{equation}
Taking the second fiber derivative of $F^2$ determines the metric tensor on the fiber, $v_{\alpha\beta}(x, y) = \frac{1}{2}\partial^2 F^2 / (\partial y^\alpha \partial y^\beta) = g_{\alpha\beta}(x) + h_{\alpha\beta}(x, y)$, where the first-order perturbation term is explicitly given by~\citep{Kapsabelis2024}
\begin{equation}
h_{\alpha\beta}(x, y) = \frac{1}{L}\left(A_\beta g_{\alpha\gamma} + A_\gamma g_{\alpha\beta} + A_\alpha g_{\beta\gamma}\right)y^\gamma + \frac{1}{L^3} A_\gamma y^\gamma g_{\alpha\epsilon}g_{\beta\delta}y^\delta y^\epsilon.
\label{eq:finsler_perturbation_h}
\end{equation}
Here, the timelike norm of the velocity vector is defined as $L \equiv \sqrt{-g_{\alpha\beta}(x)y^\alpha y^\beta}$~\citep{Kapsabelis2024}. The total geometry across $T\mathcal{M}$ is then described by a Sasaki-type metric~\citep{Sasaki1958,Kapsabelis2024}:
\begin{equation}
\mathcal{G} = g_{\mu\nu}(x)\,\mathrm{d}x^\mu \otimes \mathrm{d}x^\nu + v_{\alpha\beta}(x, y)\,\delta y^\alpha \otimes \delta y^\beta,
\label{eq:sasaki_metric}
\end{equation}
decomposed on the adapted basis $\delta y^\alpha = \mathrm{d}y^\alpha + N^\alpha_{\ \mu}\,\mathrm{d}x^\mu$ via the Cartan non-linear connection coefficients $N^\alpha_{\ \mu}(x, y)$~\citep{Kapsabelis2024}.

In applying this structure to a cosmological context, we take the base spacetime to be a flat ($k = 0$) Friedmann--Lema\^{i}tre--Robertson--Walker (FLRW) metric in spherical coordinates, $g_{\mu\nu}(x) = \mathrm{diag}\left(-1, a^2(t), a^2(t)r^2, a^2(t)r^2\sin^2\theta\right)$, with $a(t)$ denoting the cosmological scale factor~\citep{Kapsabelis2024}. The directional vectors are chosen according to the ans\"atze~\citep{Kapsabelis2024}
\begin{equation}
A_\alpha = \left(A_0(t), A_1(t), 0, 0\right), \quad y^\alpha = \left(y^0, y^1, 0, 0\right),
\label{eq:ansatze_vectors}
\end{equation}
where $y^0, y^1$ are constants and $A_0(t), A_1(t)$ are time-dependent scalar functions. Their inner products define the scalar profiles $W_0(t) \equiv A_0(t)y^0$ and $W_1(t) \equiv A_1(t)y^1$, which capture the temporal and spatial components of the geometric anisotropy, respectively~\citep{Kapsabelis2024}. Evaluating the fiber norm over this background gives
\begin{equation}
L(t) = \sqrt{(y^0)^2 - a^2(t)(y^1)^2}.
\label{eq:norm_L_explicit}
\end{equation}
To preserve reality and positivity of $L(t)$ across cosmic time, one requires the spatial velocity component to remain small, $y^1 \ll y^0$~\citep{Kapsabelis2024}.

The horizontal field equations derived from the tangent-bundle Einstein--Hilbert action read~\citep{Kapsabelis2024}
\begin{equation}
R_{\mu\nu} - \frac{1}{2}(R + S)g_{\mu\nu} + \mathcal{B}_{(\mu\nu)} - g_{\mu\nu}\mathcal{B} = \kappa T_{\mu\nu},
\label{eq:field_equations}
\end{equation}
where $\kappa \equiv 8\pi G$, $\mathcal{B}_{\mu\nu} \equiv D_\mu \mathcal{T}^\beta_{\ \nu\beta}$ is constructed from the contraction of the torsion tensor $\mathcal{T}^\beta_{\ \nu\beta}$, and $\mathcal{B} \equiv g^{\mu\nu}\mathcal{B}_{\mu\nu}$~\citep{Kapsabelis2024}. For a universe filled with pressureless dust, $T^\mu_{\ \nu} = \mathrm{diag}(-\rho_m, 0, 0, 0)$, we enforce the strictly isotropic limit by assuming a vanishing spatial fiber velocity, $y^1 \to 0$ (with $y^1 \ll y^0$). Physically, this assumption aligns the fiber velocity with a comoving cosmic observer, suppressing spatial shear while ensuring a stable, positive timelike norm $L(t) \to y^0$:
\begin{equation}
W_1(t) = 0, \quad L(t) = y^0.
\label{eq:isotropic_limit}
\end{equation}
In this limit, the energy-momentum conservation equation decouples from geometric shear, yielding the standard matter dilution law:
\begin{equation}
\dot{\rho}_m + 3H\rho_m = 0 \implies \rho_m(a) = \rho_{m0}a^{-3}.
\label{eq:continuity_equation}
\end{equation}
Inserting Eq.~\eqref{eq:isotropic_limit} into the horizontal field equations and using $\dot{W}_0 = aH(a)(\mathrm{d}W_0/\mathrm{d}a)$, the first modified Friedmann equation becomes
\begin{equation}
H^2(a)\left[1 - \frac{5}{2y^0}a\frac{\mathrm{d}W_0}{\mathrm{d}a}\right] = \frac{\kappa}{3}\rho_{m0}a^{-3},
\label{eq:friedmann_first}
\end{equation}
which can be expressed in terms of the normalized Hubble rate $E(a) \equiv H(a)/H_0$ as
\begin{equation}
E^2(a)\left[1 - \frac{5}{2y^0}a\frac{\mathrm{d}W_0}{\mathrm{d}a}\right] = \Omega_{m0}a^{-3},
\label{eq:normalized_friedmann}
\end{equation}
with $\Omega_{m0} \equiv \kappa\rho_{m0}/(3H_0^2)$. 

Since $E(1) = 1$ at the current epoch ($a = 1$), evaluating Eq.~\eqref{eq:normalized_friedmann} yields the normalization condition:
\begin{equation}
\left.\frac{5}{2y^0}\frac{\mathrm{d}W_0}{\mathrm{d}a}\right|_{a=1} = 1 - \Omega_{m0}.
\label{eq:boundary_normalization}
\end{equation}
Rather than constraining the physical matter density $\Omega_{m0}$, Eq.~\eqref{eq:boundary_normalization} fixes the characteristic scale parameter of the chosen ansatz $W_0(a)$, ensuring the model reproduces the present-day cosmic expansion without introducing unconstrained offsets.

\subsection{Effective Dark Energy Dynamics and Cosmological Diagnostics}
\label{subsec:effective_de_cosmography}

To interpret the geometric modifications in standard gravitational language, we recast the isotropic FRS Friedmann system ($p_m = 0$) into effective Einsteinian form~\citep{Kapsabelis2024}:
\begin{equation}
3H^2 = \kappa\left(\rho_m + \rho_{\mathrm{DE}}\right), \quad 2\dot{H} + 3H^2 = -\kappa p_{\mathrm{DE}}.
\label{eq:effective_einstein_equations}
\end{equation}
Comparing Eq.~\eqref{eq:effective_einstein_equations} with the reduced background relation reveals that the tangent-bundle geometry acts as an effective dark energy fluid with density and pressure:

\begin{equation}
\rho_{\mathrm{DE}}(a)
= \frac{3H_0^2}{\kappa}E^2(a)\mathcal{G}(a),
\label{eq:rho_de}
\end{equation}
where the dynamic geometric factor reads
\begin{equation}
\mathcal{G}(a) \equiv \frac{5}{2y^0}a\frac{\mathrm{d}W_0}{\mathrm{d}a}, \quad \text{with} \quad \mathcal{G}(1) = 1 - \Omega_{m0}.
\label{eq:geometric_factor_G}
\end{equation}
The corresponding fractional energy densities are $\Omega_m(a) = 1 - \mathcal{G}(a)$ and $\Omega_{\mathrm{DE}}(a) = \mathcal{G}(a)$, which identically satisfy the spatial flatness condition $\Omega_m(a) + \Omega_{\mathrm{DE}}(a) = 1$ throughout cosmic history.

Taking the ratio $w_{\mathrm{DE}} \equiv p_{\mathrm{DE}}/\rho_{\mathrm{DE}}$ provides the effective dark-energy equation-of-state parameter~\citep{Kapsabelis2024}:
\begin{equation}
w_{\mathrm{DE}}(a)
=
-1-\frac{1}{3}
\frac{\mathrm{d}\ln\rho_{\mathrm{DE}}(a)}
{\mathrm{d}\ln a}.
\label{eq:w_de_rho}
\end{equation}
Changing variables to redshift, $w_{\mathrm{DE}}$ takes the operational form:
\begin{equation}
w_{\mathrm{DE}}(z) = -1 + \frac{2(1+z)E(z)E'(z) - 3\Omega_{m0}(1+z)^3}{3\left[E^2(z) - \Omega_{m0}(1+z)^3\right]}.
\label{eq:w_de_z}
\end{equation}

We complement this dynamical description with a model-independent kinematic analysis by expanding the scale factor $a(t)$ in a Taylor series about today's epoch $t_0$~\citep{Visser2004}:
\begin{equation}
\frac{a(t)}{a(t_0)} = 1 + H_0(t - t_0) - \frac{q_0}{2!} H_0^2(t - t_0)^2 + \frac{j_0}{3!} H_0^3(t - t_0)^3 + \frac{s_0}{4!} H_0^4(t - t_0)^4 + \frac{l_0}{5!} H_0^5(t - t_0)^5 + \dots
\label{eq:taylor_scale_factor}
\end{equation}
The expansion history is characterized by the dimensionless cosmographic coefficients---deceleration ($q$), jerk ($j$), snap ($s$), and lerk ($l$)~\citep{Dunajski2008,Capozziello2019}:
\begin{equation}
q \equiv -\frac{1}{aH^2}\frac{\mathrm{d}^2 a}{\mathrm{d}t^2}, \quad j \equiv \frac{1}{aH^3}\frac{\mathrm{d}^3 a}{\mathrm{d}t^3}, \quad s \equiv \frac{1}{aH^4}\frac{\mathrm{d}^4 a}{\mathrm{d}t^4}, \quad l \equiv \frac{1}{aH^5}\frac{\mathrm{d}^5 a}{\mathrm{d}t^5}.
\label{eq:cosmographic_definitions}
\end{equation}
Applying the coordinate relation $\mathrm{d}/\mathrm{d}t = -(1+z)H(z)\mathrm{d}/\mathrm{d}z$, these parameters become explicit functions of $E(z)$ and its redshift derivatives:
\begin{align}
q(z) &= -1 + (1+z)\frac{E'(z)}{E(z)}, \label{eq:q_z} \\
j(z) &= 1 - 2(1+z)\frac{E'(z)}{E(z)} + (1+z)^2\left[\frac{E''(z)}{E(z)} + \left(\frac{E'(z)}{E(z)}\right)^2\right], \label{eq:j_z} \\
s(z) &= -\left[2 + 3q(z)\right]j(z) - (1+z)j'(z), \label{eq:s_z} \\
l(z) &= -\left[3 + 4q(z)\right]s(z) - (1+z)s'(z). \label{eq:l_z}
\end{align}
In a spatially flat $\Lambda\mathrm{CDM}$ universe, the jerk parameter remains strictly pinned at unity ($j_{\Lambda\mathrm{CDM}} = 1$) for all epochs~\citep{Sahni2003,Alam2003}, forcing the snap parameter to follow $s_{\Lambda\mathrm{CDM}}(z) = 1 - \frac{9}{2}\Omega_m(z)$. Any observed evolution where $j(z) \neq 1$ thus provides a direct geometrical signal of non-$\Lambda\mathrm{CDM}$ behavior.

Finally, we track the geometric statefinder trajectory $\{r, s_{\mathrm{sf}}\}$ and the two-point $Om(z)$ diagnostic~\citep{Sahni2003,Sahni2008}:
\begin{equation}
r(z) \equiv j(z), \quad s_{\mathrm{sf}}(z) \equiv \frac{r(z) - 1}{3\left[q(z) - \frac{1}{2}\right]}, \quad Om(z) \equiv \frac{E^2(z) - 1}{(1+z)^3 - 1}.
\label{eq:diagnostics}
\end{equation}
Under standard flat $\Lambda\mathrm{CDM}$, the statefinder pair is localized at the fixed point $\{r, s_{\mathrm{sf}}\} = \{1, 0\}$, while the $Om(z)$ diagnostic collapses to the constant value $Om(z) = \Omega_{m0}$ across all redshifts. Departures from these baseline values expose the dynamical imprint of the Finslerian geometric corrections.

\section{Cosmological Parametrizations}
\label{sec:cosmological_parametrizations}

To quantitatively trace background expansion, we investigate two phenomenological forms for the temporal anisotropic scalar $W_0(a)$: a logarithmic evolution and an exponential profile. In each model, the boundary normalization condition~\eqref{eq:boundary_normalization} fixes the coupling amplitude $\alpha$ in terms of $\Omega_{m0}$, preventing unconstrained boundary offsets. The logarithmic form admits an exact $\Lambda$CDM limit, whereas the exponential form provides a complementary smooth transition, allowing the observational consequences of different geometric evolutions to be tested.

\subsection{Model I: Logarithmic Parametrization}
\label{subsec:model_1_log}

We first model the temporal anisotropy with a logarithmic dependence on the cosmic scale factor:
\begin{equation}
W_0(a) = y^0\alpha\ln\left(1 + \beta a^n\right),
\label{eq:w0_log}
\end{equation}
where $\alpha$ is the coupling strength, $\beta > 0$ sets the characteristic transition scale, and $n > 0$ governs the power-law growth rate of the tangent-bundle perturbation. Differentiating Eq.~\eqref{eq:w0_log} with respect to $a$ gives
\begin{equation}
\frac{\mathrm{d}W_0}{\mathrm{d}a} = y^0\alpha\frac{n\beta a^{n-1}}{1 + \beta a^n}.
\label{eq:dw0_da_log}
\end{equation}
Enforcing the normalization constraint~\eqref{eq:boundary_normalization} at $a = 1$ uniquely determines the amplitude:
\begin{equation}
\alpha = \frac{2(1 - \Omega_{m0})(1 + \beta)}{5n\beta}.
\label{eq:alpha_log}
\end{equation}
Substituting Eq.~\eqref{eq:alpha_log} into the dynamic geometric function $\mathcal{G}(a) \equiv \frac{5}{2y^0}a\frac{\mathrm{d}W_0}{\mathrm{d}a}$ yields
\begin{equation}
\mathcal{G}_{\log}(a) = (1 - \Omega_{m0})a^n\left(\frac{1 + \beta}{1 + \beta a^n}\right).
\label{eq:G_log_a}
\end{equation}
Inserting $\mathcal{G}_{\log}(a)$ into the horizontal Friedmann relation $E^2(a)[1 - \mathcal{G}(a)] = \Omega_{m0}a^{-3}$ gives the normalized expansion rate:
\begin{equation}
E^2_{\log}(a) = \frac{\Omega_{m0}a^{-3}}{1 - (1 - \Omega_{m0})a^n\left(\frac{1+\beta}{1+\beta a^n}\right)}.
\label{eq:E2_log_a}
\end{equation}
In redshift space ($a = (1+z)^{-1}$), this becomes
\begin{equation}
E^2_{\log}(z) = \frac{\Omega_{m0}(1 + z)^3}{1 - \frac{(1 - \Omega_{m0})(1 + \beta)}{(1 + z)^n + \beta}}.
\label{eq:E2_log_z}
\end{equation}
Setting $n = 3$ and $\beta = (1 - \Omega_{m0})/\Omega_{m0}$ reduces Eq.~\eqref{eq:E2_log_z} analytically to the flat $\Lambda\mathrm{CDM}$ background,
\begin{equation}
E^2(z) = \Omega_{m0}(1 + z)^3 + 1 - \Omega_{m0},
\label{eq:E2_lcdm}
\end{equation}
At present epoch Eq.~\eqref{eq:w_de_z} yields the current effective dark-energy equation of state:
\begin{equation}
w_0^{\log} \equiv w_{\mathrm{DE}}(0) = -\frac{n}{3\Omega_{m0}(1 + \beta)},
\label{eq:w0_log1}
\end{equation}
across all redshifts, with $w_0^{\log} = -1$ a. More generally, the current dark energy behaves as quintessence ($w_0^{\log} > -1$) when $n < 3\Omega_{m0}(1 + \beta)$ and crosses into the phantom regime ($w_0^{\log} < -1$) when $n > 3\Omega_{m0}(1 + \beta)$.

\subsection{Model II: Exponential Parametrization}
\label{subsec:model_2_exp}

Our second ansatz assumes an exponential form for the temporal anisotropy:
\begin{equation}
W_0(a) = y^0\alpha\left(1 - \mathrm{e}^{-pa^2}\right),
\label{eq:w0_exp}
\end{equation}
where $\alpha$ is the coupling amplitude and the dimensionless factor $p$ controls the transition rate between epochs. Differentiating Eq.~\eqref{eq:w0_exp} yields
\begin{equation}
\frac{\mathrm{d}W_0}{\mathrm{d}a} = 2y^0\alpha p a\,\mathrm{e}^{-pa^2}.
\label{eq:dw0_da_exp}
\end{equation}
Applying the boundary condition~\eqref{eq:boundary_normalization} at $a = 1$ locks the amplitude:
\begin{equation}
\alpha = \frac{1 - \Omega_{m0}}{5p\,\mathrm{e}^{-p}}.
\label{eq:alpha_exp}
\end{equation}
The resulting geometric factor is
\begin{equation}
\mathcal{G}_{\exp}(a) = (1 - \Omega_{m0})a^2\mathrm{e}^{-p(a^2 - 1)}.
\label{eq:G_exp_a}
\end{equation}
The normalized expansion rate as a function of scale factor reads
\begin{equation}
E^2_{\exp}(a) = \frac{\Omega_{m0}a^{-3}}{1 - (1 - \Omega_{m0})a^2\mathrm{e}^{-p(a^2 - 1)}},
\label{eq:E2_exp_a}
\end{equation}
which in redshift space evaluates to
\begin{equation}
E^2_{\exp}(z) = \frac{\Omega_{m0}(1 + z)^3}{1 - \frac{1 - \Omega_{m0}}{(1 + z)^2}\exp\left[-p\left((1 + z)^{-2} - 1\right)\right]}.
\label{eq:E2_exp_z}
\end{equation}
Evaluating the effective dark-energy equation of state today yields
\begin{equation}
w_0^{\exp} \equiv w_{\mathrm{DE}}(0) = -\frac{2(1 - p)}{3\Omega_{m0}}.
\label{eq:w0_exp1}
\end{equation}
Imposing $w_0^{\exp} = -1$ leads to the parameter condition
\begin{equation}
p = 1 - \frac{3}{2}\Omega_{m0}.
\label{eq:p_lcdm_condition}
\end{equation}
This condition matches the effective dark-energy equation of state to that of a cosmological constant only at the current epoch ($z = 0$). Unlike Logarithmic Parametrization, Exponential Parametrization cannot reproduce $\Lambda\mathrm{CDM}$ globally because the transcendental factor $\mathcal{G}_{\exp}(a) = (1 - \Omega_{m0})a^2\mathrm{e}^{-p(a^2 - 1)}$ cannot match the algebraic form $\mathcal{G}_{\Lambda\mathrm{CDM}}(a) = (1 - \Omega_{m0})/[\Omega_{m0}a^{-3} + 1 - \Omega_{m0}]$ across all $a$ for any constant $p$. At the present epoch, the dark energy behaves as quintessence ($w_0^{\exp} > -1$) for $p > 1 - \frac{3}{2}\Omega_{m0}$ and enters the phantom domain ($w_0^{\exp} < -1$) for $p < 1 - \frac{3}{2}\Omega_{m0}$.

\section{Observational Data and Statistical Methodology}
\label{sec:data}

To constrain the background dynamics of the Finsler--Randers--Sasaki (FRS) framework, we perform a Bayesian Markov Chain Monte Carlo (MCMC) analysis using the affine-invariant ensemble sampler \texttt{emcee}~\citep{foreman2013emcee}. We investigate two observational configurations to quantify the sensitivity of the Finslerian parameters to local distance anchors:
\begin{itemize}
    \item \textbf{CC + BAO + PP:} The baseline dataset, combining 30 cosmic chronometer expansion rates, 13 entries in the DESI DR2 BAO data vector and 1580 Pantheon+ Hubble-flow supernovae with an analytically marginalized absolute magnitude $M$.
    \item \textbf{CC + BAO + PPS:} The baseline suite augmented by the local distance-ladder anchor from SH0ES, incorporated as an explicit Gaussian prior on the Hubble constant, $H_0 = 73.04 \pm 1.04\ \mathrm{km\,s^{-1}\,Mpc^{-1}}$~\citep{Riess2022}.
\end{itemize}

\subsection{Cosmic Chronometers (CC)} 
\label{subsec:cc_data}

Cosmic Chronometers yield direct determinations of the Hubble expansion rate $H(z)$ that avoid adopting an assumed fiducial cosmology or performing line-of-sight metric integrations~\citep{Jimenez2002}. By examining pairs of massive, passively evolving early-type galaxies with comparable chemical abundances across small redshift separations $\Delta z$, the expansion rate is obtained from the differential age relation:
\begin{equation}
    H(z) = -\frac{1}{1+z}\,\frac{\mathrm{d}z}{\mathrm{d}t} \approx -\frac{1}{1+z}\,\frac{\Delta z}{\Delta t},
    \label{eq:cc_relation}
\end{equation}
where age differences $\Delta t$ are extracted spectroscopically from age-sensitive features such as the $D_n4000$ break~\citep{Moresco:2012jh,Moresco:2016mzx}.

We adopt an updated compilation of 30 $H(z)$ measurements spanning $0.070 \le z \le 1.750$~\citep{Moresco:2016mzx,Moresco:2020lyq,Moresco:2022phi}. For this catalog, the likelihood contribution is evaluated using a diagonal covariance:
\begin{equation}
    \chi^2_{\mathrm{CC}} = \sum_{i=1}^{30} \left[ \frac{H_0 E(z_i) - H_{\mathrm{obs}}(z_i)}{\sigma_{H,i}} \right]^2,
    \label{eq:chi2_cc}
\end{equation}
where $\sigma_{H,i}$ accounts for statistical measurement errors and systematic uncertainties associated with stellar population synthesis libraries, metallicity spreads, and star formation histories~\citep{Moresco:2020lyq}.

\subsection{DESI DR2 BAO Constraints}
\label{subsec:desi_bao}

Baryon Acoustic Oscillations (BAO) provide an uncalibrated standard ruler set by the sound horizon at the drag epoch, $r_d$. In this analysis, we implement the BAO likelihood from the Dark Energy Spectroscopic Instrument (DESI) Data Release 2 (DR2)~\citep{DESI:2025DR2II}. The dataset covers the effective redshift range $0.295 \le z_{\rm eff} \le 2.330$~\citep{DESI:2025DR2II,AbdulKarim:2025dr2i}. The observable distances depend on the expansion rate $H(z)$ along the line of sight via the Hubble distance $D_H(z) \equiv c/H(z)$, and across the transverse plane via $D_M(z) \equiv c\int_0^z \mathrm{d}z'/H(z')$. For isotropic clustering measurements, these project into the spherically averaged distance scale:
\begin{equation}
D_V(z) \equiv \left[ z D_M^2(z) D_H(z) \right]^{1/3}.
\label{eq:dv_def}
\end{equation}
The DESI DR2 data vector $\mathbf{y}_{\mathrm{obs}}$ comprises 13 points: an isotropic ratio $D_V/r_d$ at $z_{\mathrm{eff}}=0.295$ from the Bright Galaxy Sample, followed by six anisotropic pairs of $D_M/r_d$ and $D_H/r_d$. These pairs correspond to LRGs ($z_{\mathrm{eff}} \in \{0.510, 0.706\}$), the joint LRG+ELG sample ($z_{\mathrm{eff}}=0.934$), ELGs ($z_{\mathrm{eff}}=1.321$), quasars ($z_{\mathrm{eff}}=1.484$), and the Lyman-$\alpha$ forest ($z_{\mathrm{eff}}=2.330$). For Lyman-$\alpha$, the published vector orders the entries as $(D_H/r_d, D_M/r_d)$~\citep{AbdulKarim:2025dr2i}, which we match in our likelihood calculation. We evaluate the BAO likelihood via the standard $\chi^2$ statistic:
\begin{equation}
\chi^2_{\mathrm{BAO}}(\Theta) =
\left[\mathbf{y}_{\mathrm{obs}}-\mathbf{y}_{\mathrm{th}}(\Theta)\right]^T
\mathbf{C}_{\mathrm{BAO}}^{-1}
\left[\mathbf{y}_{\mathrm{obs}}-\mathbf{y}_{\mathrm{th}}(\Theta)\right],
\label{eq:chi2_bao}
\end{equation}
where $\mathbf{C}_{\mathrm{BAO}}$ denotes the full $13\times13$ covariance matrix from the DESI DR2 release.

\subsection{Type Ia Supernovae Compilation and Likelihood Formulations}
\label{subsec:pantheon_data}

Luminosity distances are calibrated against the Pantheon+ Type~Ia supernova (SN~Ia) compilation, which aggregates 1701 light curves corresponding to 1550 spectroscopically confirmed events across the redshift baseline $0.001 \le z \le 2.26$~\citep{Scolnic:2022pnt,Brout:2022cos}. The theoretical distance modulus relates to the transverse comoving distance $D_M(z)$ via the luminosity distance $D_L(z) = (1+z)D_M(z)$:
\begin{equation}
    \mu_{\mathrm{th}}(z_i, \Theta) = 5\log_{10}\left[\frac{D_L(z_i)}{1\ \mathrm{Mpc}}\right] + 25.
    \label{eq:mu_th}
\end{equation}
For an observed apparent $B$-band maximum magnitude $m_{B,i}$ and fiducial absolute magnitude $M$, the observational distance modulus is $\mu_i = m_{B,i} - M$. The corresponding chi-square statistic takes the quadratic form:
\begin{equation}
    \chi^2_{\mathrm{SN}} = \Delta \boldsymbol{\mu}^T \mathbf{C}_{\mathrm{SN}}^{-1} \Delta \boldsymbol{\mu},
    \label{eq:chi2_sn}
\end{equation}
where $\Delta \boldsymbol{\mu} \equiv \mathbf{m}_B - M\mathbf{1} - \boldsymbol{\mu}_{\mathrm{th}}$, and $\mathbf{C}_{\mathrm{SN}}$ denotes the complete covariance matrix encompassing both observational statistical noise and systematic calibration uncertainties~\citep{Scolnic:2022pnt}.

We investigate two distinct operational treatments of this compilation to assess the influence of local distance anchors on the underlying Finslerian parameters:
\begin{itemize}
    \item \textbf{PP (Uncalibrated Pantheon+):} To decouple the background expansion from local peculiar velocity dispersions and volumetric selection effects, we discard the 77 Cepheid-host calibrators ($\mathtt{IS\_CALIBRATOR} = 0$) and restrict the Hubble-flow sample to $z_{\mathrm{HD}} > 0.01$. This cut leaves $N_{\mathrm{SN}} = 1580$ supernova light curves, with the covariance matrix reduced accordingly. In the absence of an absolute distance ladder, an exact degeneracy persists between $H_0$ and $M$. We bypass this degenerate degree of freedom by marginalizing over $M$ analytically under an improper uniform prior.

    \item \textbf{PPS (Pantheon+ with SH0ES Anchor):} In this second configuration, the 1580 Hubble-flow supernovae are evaluated alongside the local distance-ladder calibration from the SH0ES program~\citep{Riess2022}. Rather than refitting individual host Cepheid light curves, this constraint enters as an independent Gaussian prior on the Hubble constant, $H_0 = 73.04 \pm 1.04\ \mathrm{km\,s^{-1}\,Mpc^{-1}}$, added to the total log-likelihood as $\chi^2_{\mathrm{SH0ES}} = \left[(H_0 - 73.04)/1.04\right]^2$. This anchor breaks the structural $H_0$--$M$ degeneracy, fixing the absolute magnitude scale $M$ and enabling tight, localized constraints on both $H_0$ and the Finslerian transition parameter $p$.
\end{itemize}

\section{Results and Discussion}
\label{sec:results_discussion}

\subsection{Background Expansion and Parameter Estimation}
\label{subsec:background_dynamics}

\begin{figure}[H]
\centering

\begin{minipage}{0.48\textwidth}
\centering
\includegraphics[width=\linewidth]{
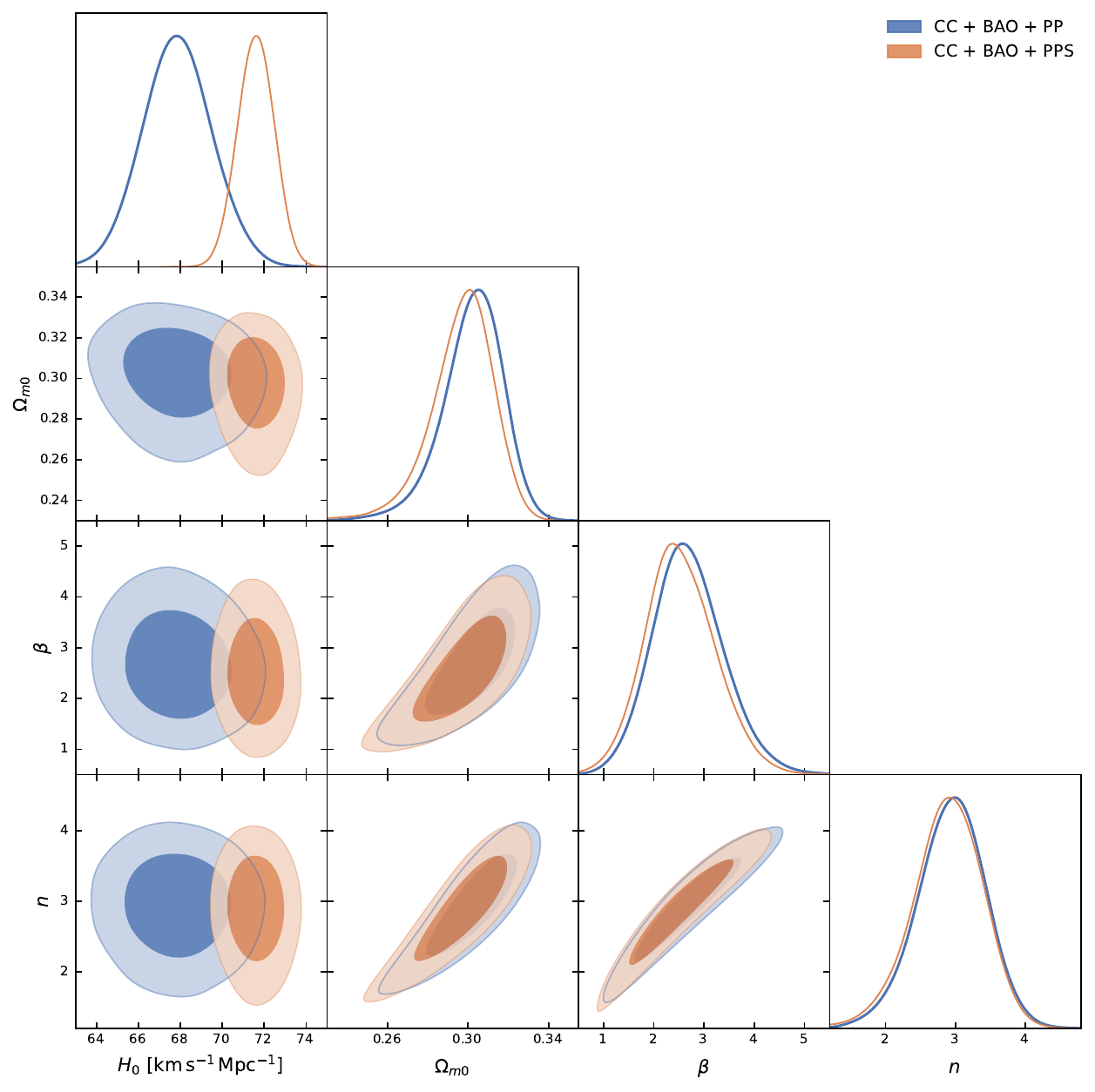
}

\vspace{2pt}

\small (a) FRS Logarithmic model
\end{minipage}
\hfill
\begin{minipage}{0.48\textwidth}
\centering
\includegraphics[width=\linewidth]{
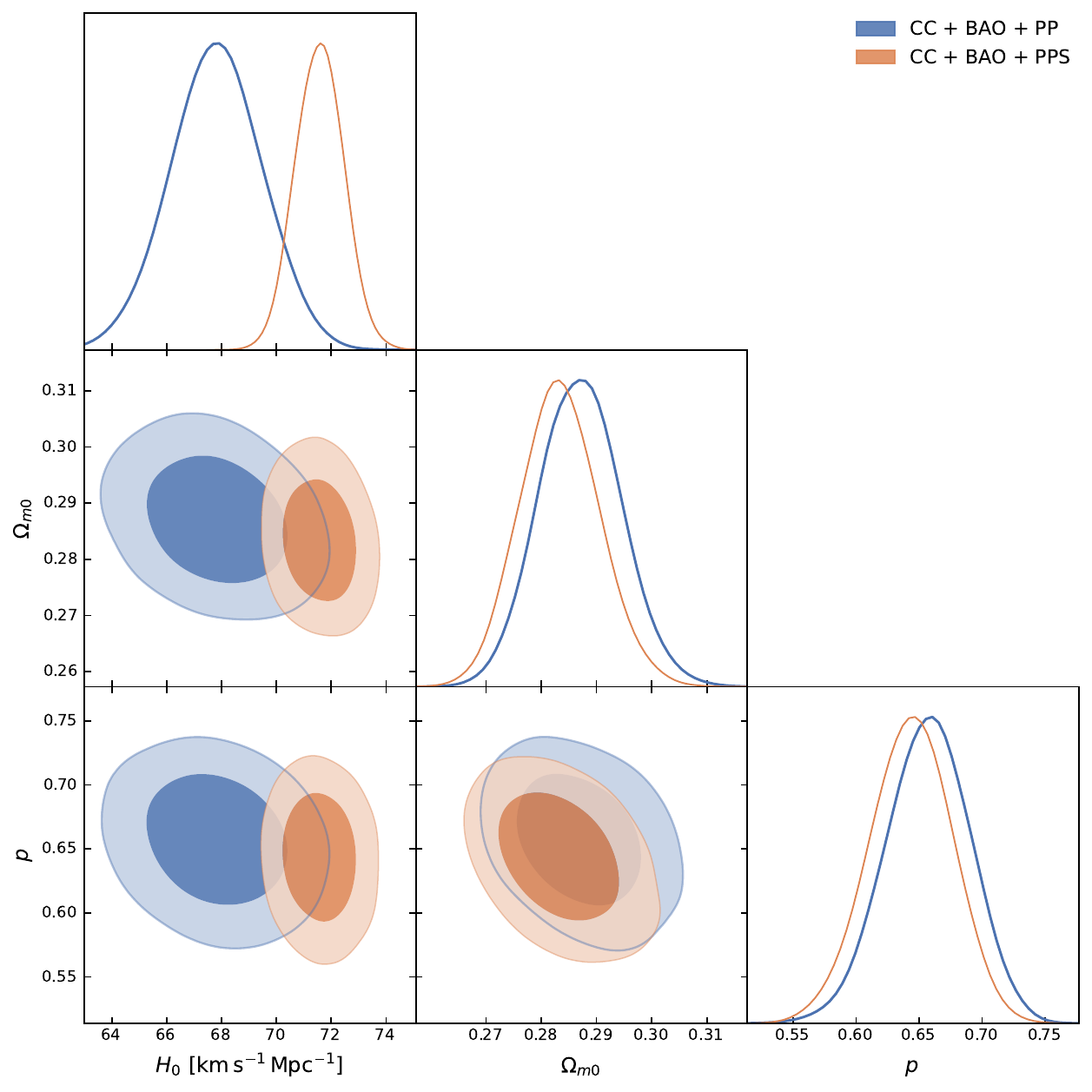
}

\vspace{2pt}

\small (b) FRS Exponential model
\end{minipage}

\caption{Marginalized one-dimensional posterior distributions
and two-dimensional $68\%$ and $95\%$ confidence contours for the
$\mathrm{FRS}_{\mathrm{LOG}}$ and
$\mathrm{FRS}_{\mathrm{EXP}}$ gravity models. Constraints are
derived from the background observational dataset combinations
CC+BAO+PP and CC+BAO+PPS.}

\label{fig:combined_mcmc_contours}

\end{figure}

The joint observational constraints on flat $\Lambda\mathrm{CDM}$ and the two Finsler--Randers--Sasaki formulations ($\mathrm{FRS}_{\mathrm{LOG}}$ and $\mathrm{FRS}_{\mathrm{EXP}}$) are compiled in Table~\ref{tab:cosmological_parameters_comparison}.The corresponding one-dimensional posterior distributions and two-dimensional joint confidence contours ($68\%$ and $95\%$ credible regions) are illustrated in Fig.~\ref{fig:combined_mcmc_contours}

Under the baseline compilation ($\mathrm{CC}+\mathrm{BAO}+\mathrm{PP}$), both FRS geometries infer expansion rates that are mutually consistent within $1\sigma$: $H_0 = 67.86 \pm 1.60\ \mathrm{km\,s^{-1}\,Mpc^{-1}}$ for the logarithmic profile and $H_0 = 67.82 \pm 1.70\ \mathrm{km\,s^{-1}\,Mpc^{-1}}$ for the exponential profile. Both values settle slightly lower than the reference flat $\Lambda\mathrm{CDM}$ estimate of $H_0 = 68.85_{-1.63}^{+1.62}\ \mathrm{km\,s^{-1}\,Mpc^{-1}}$. For $\mathrm{FRS}_{\mathrm{LOG}}$, the matter density parameter $\Omega_{m0} = 0.3020_{-0.0120}^{+0.0170}$ tracks the standard cosmological value ($\Omega_{m0} = 0.3049_{-0.0077}^{+0.0079}$) closely, while the deformation exponents converge toward $\beta = 2.69_{-0.77}^{+0.64}$ and $n = 2.94_{-0.45}^{+0.51}$. Because $n$ remains statistically consistent with $3$, the logarithmic ansatz comfortably accommodates its global $\Lambda\mathrm{CDM}$ limit. In contrast, the exponential parametrization favors a lower matter density, $\Omega_{m0} = 0.2870 \pm 0.0074$, offset by a non-zero Finslerian coupling parameter $p = 0.657_{-0.032}^{+0.036}$.
The sound horizon at the drag epoch also remains remarkably steady across all three setups. We find $r_{\mathrm{d}} = 146.63^{+3.48}_{-3.32}\,\mathrm{Mpc}$ for flat $\Lambda\mathrm{CDM}$, matching the shared estimate of $r_{\mathrm{d}} = 147.0 \pm 3.4\,\mathrm{Mpc}$ recovered by both Finsler--Randers profiles.

Introducing the SH0ES Cepheid distance-ladder anchor ($\mathrm{CC}+\mathrm{BAO}+\mathrm{PPS}$) shifts the posterior distributions toward higher expansion rates across all three models. The Hubble constant increases to $H_0 = 71.85_{-0.88}^{+0.87}\ \mathrm{km\,s^{-1}\,Mpc^{-1}}$ in $\Lambda\mathrm{CDM}$, while reaching $H_0 = 71.61 \pm 0.87\ \mathrm{km\,s^{-1}\,Mpc^{-1}}$ and $H_0 = 71.59 \pm 0.90\ \mathrm{km\,s^{-1}\,Mpc^{-1}}$ for $\mathrm{FRS}_{\mathrm{LOG}}$ and $\mathrm{FRS}_{\mathrm{EXP}}$, respectively. This systematic upward shift of $\sim 5.5\%$ is accompanied by a slight drop in the inferred matter density ($\Omega_{m0} = 0.2970_{-0.0120}^{+0.0170}$ for $\mathrm{FRS}_{\mathrm{LOG}}$ and $\Omega_{m0} = 0.2834 \pm 0.0071$ for $\mathrm{FRS}_{\mathrm{EXP}}$), reflecting the anti-correlation between $H_0$ and $\Omega_{m0}$ driven by uncalibrated supernova distance moduli. Crucially, the non-Riemannian geometric parameters remain stable under the addition of the SH0ES calibration. The logarithmic exponents stay within $1\sigma$ of their baseline values ($\beta = 2.54_{-0.72}^{+0.64}$, $n = 2.89_{-0.45}^{+0.53}$), and the exponential coupling exhibits minimal drift ($p = 0.643 \pm 0.033$). The inferred sound horizon at the drag epoch consistently settles at $r_d = 140.10 \pm 1.90\ \mathrm{Mpc}$ for both FRS models, compared to $r_d = 140.90_{-1.82}^{+1.87}\ \mathrm{Mpc}$ for $\Lambda\mathrm{CDM}$. Concurrently, the reconstructed supernova absolute magnitude anchors tightly at $M \approx -19.30\ \mathrm{mag}$, confirming that the geometric modifications operate without distorting standard low-redshift standard-candle calibrations.

\begin{table}[H]
\centering

\caption{Marginalized best-fit values with $68\%$ confidence intervals for
cosmological parameters ($H_0$, $\Omega_{m0}$, $\beta$, $n$, $p$,
$r_d$, and $M$) constrained by observational dataset
combinations under flat $\Lambda\mathrm{CDM}$, $\mathrm{FRS}_{\mathrm{LOG}}$ and
$\mathrm{FRS}_{\mathrm{EXP}}$ gravity models.}
\label{tab:cosmological_parameters_comparison}

\vspace{6pt}

\setlength{\tabcolsep}{4.5pt}
\renewcommand{\arraystretch}{1.35}

\resizebox{\textwidth}{!}{%
\begin{tabular}{llccccccc}
\toprule

Dataset & Model &
$H_0\ (\mathrm{km\ s^{-1}\ Mpc^{-1}})$ &
$\Omega_{m0}$ &
$\beta$ &
$n$ &
$p$ &
$r_d\ (\mathrm{Mpc})$ &
$M\ (\mathrm{mag})$ \\

\midrule

\multirow{3}{*}{CC+BAO+PP}
& $\Lambda\mathrm{CDM}$
& $68.85_{-1.63}^{+1.62}$
& $0.3049_{-0.0077}^{+0.0079}$
& \text{--}
& \text{--}
& \text{--}
&  $146.63_{-3.32}^{+3.48}$
& \text{--} \\

& $\mathrm{FRS}_{\mathrm{LOG}}$
& $67.86 \pm 1.60\phantom{_{-0}^{+0}}$
& $0.3020_{-0.0120}^{+0.0170}$
& $2.69_{-0.77}^{+0.64}$
& $2.94_{-0.45}^{+0.51}$
& \text{--}
& $147.0\pm 3.4\phantom{_{-0}^{+0}}$
& \text{--} \\

& $\mathrm{FRS}_{\mathrm{EXP}}$
& $67.82 \pm 1.70\phantom{_{-0}^{+0}}$
& $0.2870 \pm 0.0074\phantom{_{-0}^{+0}}$
& \text{--}
& \text{--}
& $0.657_{-0.032}^{+0.036}$
& $147.0\pm 3.4\phantom{_{-0}^{+0}}$
& \text{--} \\

\midrule

\multirow{3}{*}{CC+BAO+PPS}
& $\Lambda\mathrm{CDM}$
& $71.85_{-0.88}^{+0.87}$
& $0.3015_{-0.0076}^{+0.0078}$
& \text{--}
& \text{--}
& \text{--}
& $140.90_{-1.82}^{+1.87}$
& $-19.305_{-0.027}^{+0.026}$ \\

& $\mathrm{FRS}_{\mathrm{LOG}}$
& $71.61 \pm 0.87\phantom{_{-0}^{+0}}$
& $0.2970_{-0.0120}^{+0.0170}$
& $2.54_{-0.72}^{+0.64}$
& $2.89_{-0.45}^{+0.53}$
& \text{--}
& $140.10 \pm 1.90\phantom{_{-0}^{+0}}$
& $-19.302 \pm 0.027\phantom{_{-0}^{+0}}$ \\

& $\mathrm{FRS}_{\mathrm{EXP}}$
& $71.59 \pm 0.90\phantom{_{-0}^{+0}}$
& $0.2834 \pm 0.0071\phantom{_{-0}^{+0}}$
& \text{--}
& \text{--}
& $0.643 \pm 0.033\phantom{_{-0}^{+0}}$
& $140.10 \pm 1.90\phantom{_{-0}^{+0}}$
& $-19.301 \pm 0.027\phantom{_{-0}^{+0}}$ \\

\bottomrule
\end{tabular}%
}

\end{table}

\noindent

\subsection{Comparison with Contemporary Observations}
\label{sec:comparison_contemporary}

\begin{figure}[H]
\vspace{-0.45cm}
\centering
\includegraphics[width=\textwidth]{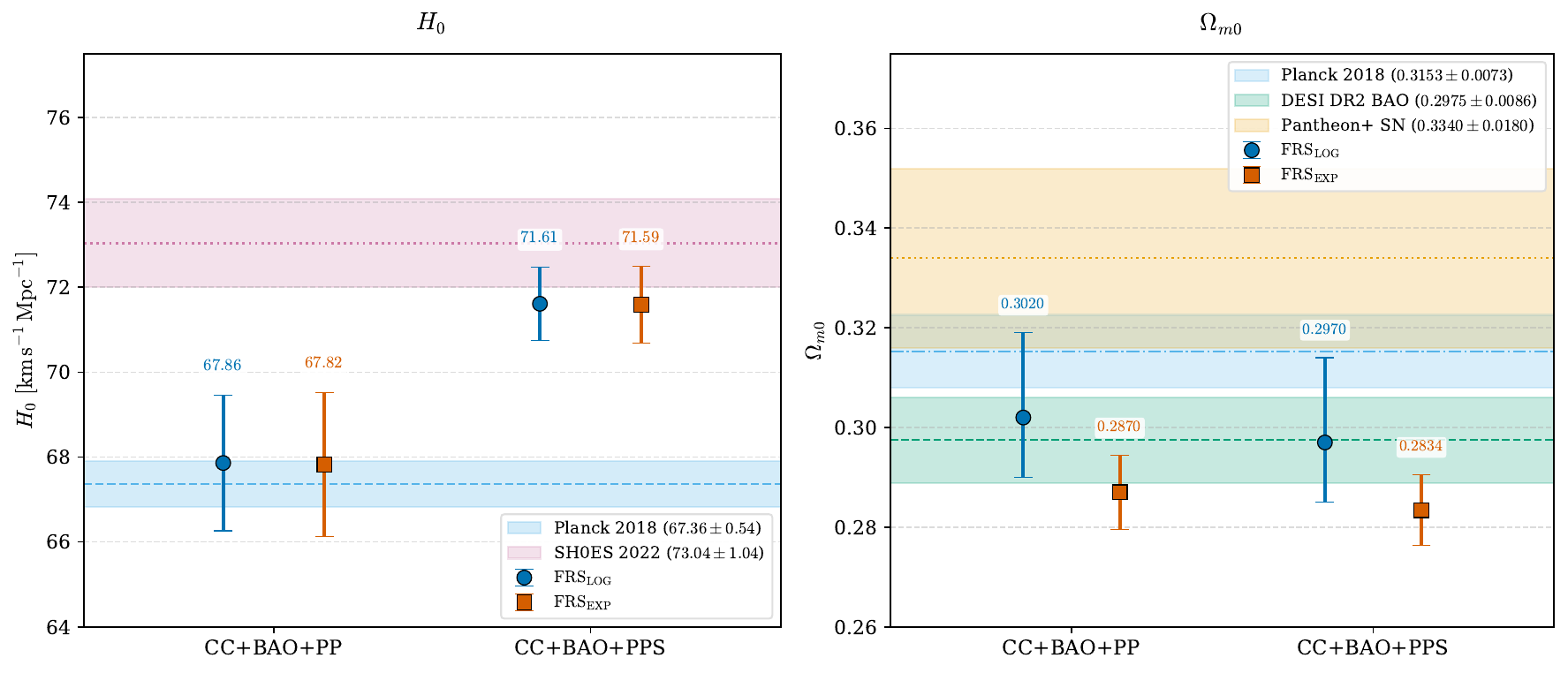}
\caption{Comparison of the marginalized constraints  obtained from  $\mathrm{FRS}_{\mathrm{LOG}}$ and
$\mathrm{FRS}_{\mathrm{EXP}}$ models across  $\mathrm{CC}+\mathrm{BAO}+\mathrm{PP}$ and $\mathrm{CC}+\mathrm{BAO}+\mathrm{PPS}$. Reference determinations from Planck~2018, SH0ES~2022, Pantheon+ and DESI~DR2.}
\label{fig:contemporary_comparison}
\end{figure}

To contextualize the observational performance of the Finsler--Randers--Sasaki frameworks, we benchmark their marginalized posterior constraints against primary cosmological probes: the Planck 2018 CMB baseline (P18)~\citep{Planck2020}, the SH0ES local distance ladder (R22)~\citep{Riess2022}, the Pantheon+ Type~Ia supernova catalog~\citep{Scolnic:2022pnt,Brout:2022cos} and the DESI DR2 BAO measurements~\citep{DESI:2025DR2II}, as illustrated in Fig.~\ref{fig:contemporary_comparison}.

Under the uncalibrated baseline ($\mathrm{CC}+\mathrm{BAO}+\mathrm{PP}$), both FRS models yield expansion rates close to the Planck baseline, giving $H_0 = 67.86 \pm 1.60\ \mathrm{km\,s^{-1}\,Mpc^{-1}}$ for $\mathrm{FRS}_{\mathrm{LOG}}$ and $H_0 = 67.82 \pm 1.70\ \mathrm{km\,s^{-1}\,Mpc^{-1}}$ for $\mathrm{FRS}_{\mathrm{EXP}}$ both within $0.3\sigma$ of the Planck determination ($H_0 = 67.36 \pm 0.54\ \mathrm{km\,s^{-1}\,Mpc^{-1}}$)~\citep{Planck2020}, while remaining in $\sim 2.6\text{--}2.7\sigma$ tension with SH0ES ($H_0 = 73.04 \pm 1.04\ \mathrm{km\,s^{-1}\,Mpc^{-1}}$)~\citep{Riess2022}. The matter density constraints, however, distinguish the two frameworks: the logarithmic ansatz gives $\Omega_{m0} = 0.3020_{-0.0120}^{+0.0170}$, matching the flat-$\Lambda\mathrm{CDM}$ constraint from DESI DR2 BAO ($\Omega_{m0} = 0.2975 \pm 0.0086$)~\citep{DESI:2025DR2II} to within $0.3\sigma$, whereas the exponential model yields a lower value, $\Omega_{m0} = 0.2870 \pm 0.0074$, differing from the uncalibrated SN-only Pantheon+ baseline ($\Omega_{m0} = 0.334 \pm 0.018$)~\citep{Brout:2022cos} at the $2.4\sigma$ level. Calibrating with the Cepheid prior ($\mathrm{CC}+\mathrm{BAO}+\mathrm{PPS}$) shifts the Hubble constant upward to $H_0 = 71.61 \pm 0.87\ \mathrm{km\,s^{-1}\,Mpc^{-1}}$ ($\mathrm{FRS}_{\mathrm{LOG}}$) and $H_0 = 71.59 \pm 0.90\ \mathrm{km\,s^{-1}\,Mpc^{-1}}$ ($\mathrm{FRS}_{\mathrm{EXP}}$), easing the SH0ES discrepancy to $\sim 1.1\sigma$~\citep{Riess2022}.

\subsection{Cosmographic Parameters and Kinematics of Cosmic Expansion}
\label{sec:cosmography}

To assess cosmic acceleration without assuming a specific dark energy equation of state, we employ the kinematic series $\{q(z), j(z), s(z), l(z)\}$, which maps the dimensionless Hubble rate $E(z)$ and its redshift derivatives. We evaluate these kinematic trajectories using the best-fit $E(z)$ profiles from our $\text{CC}+\text{BAO}+\text{PP}$ and $\text{CC}+\text{BAO}+\text{PPS}$ maximum-likelihood joint fits, benchmarking them against a flat $\Lambda\text{CDM}$ baseline ($\Omega_{m0} \approx 0.30$, $j_{\Lambda\text{CDM}} \equiv 1$). The resulting present-day parameters ($q_0, j_0, s_0, l_0$) and deceleration-to-acceleration transition redshifts $z_t$ are summarized in table~\ref{tab:cosmography_summary}.

\subsubsection{Deceleration Dynamics and Transition Redshift}
\label{sec:deceleration}

\begin{figure}[H]
\centering
\begin{minipage}[t]{0.48\textwidth}
  \centering
  \includegraphics[width=\textwidth]{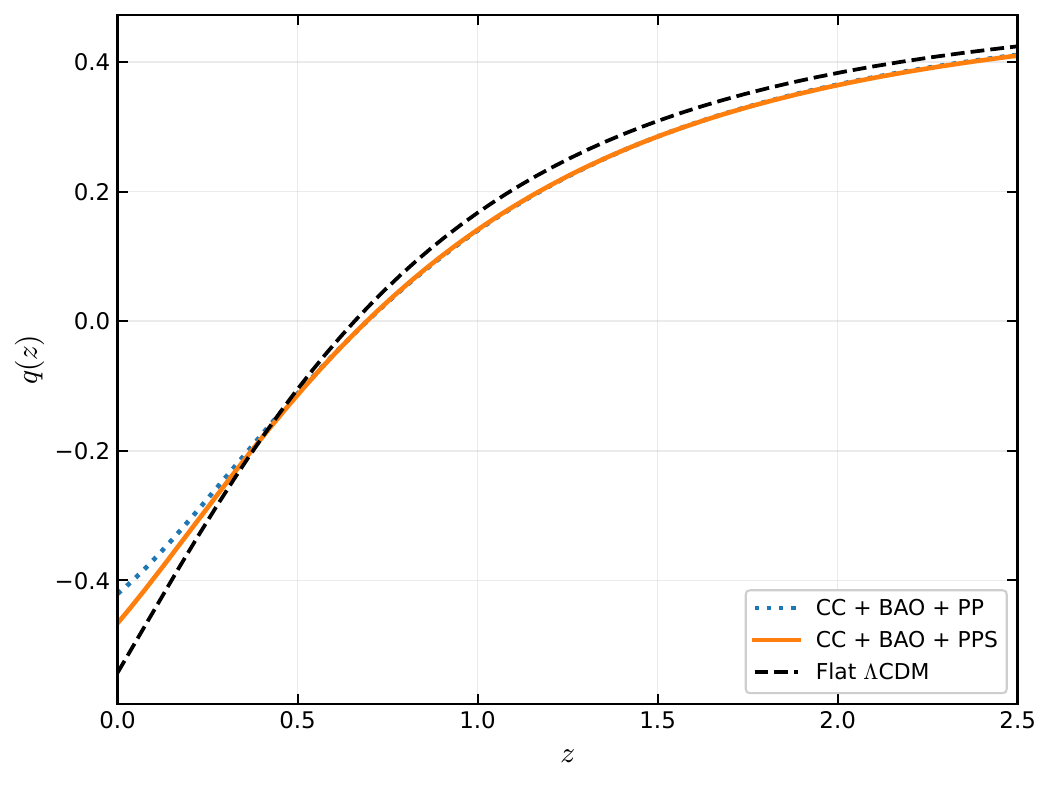}
  \caption{Deceleration parameter $q(z)$ as a function of redshift $z$ for the $\mathrm{FRS}_{\mathrm{LOG}}$ framework under $\text{CC}+\text{BAO}+\text{PP}$ and $\text{CC}+\text{BAO}+\text{PPS}$ datasets, compared to the flat $\Lambda\text{CDM}$ baseline.}
  \label{fig:q_log}
\end{minipage}
\hfill
\begin{minipage}[t]{0.48\textwidth}
  \centering
  \includegraphics[width=\textwidth]{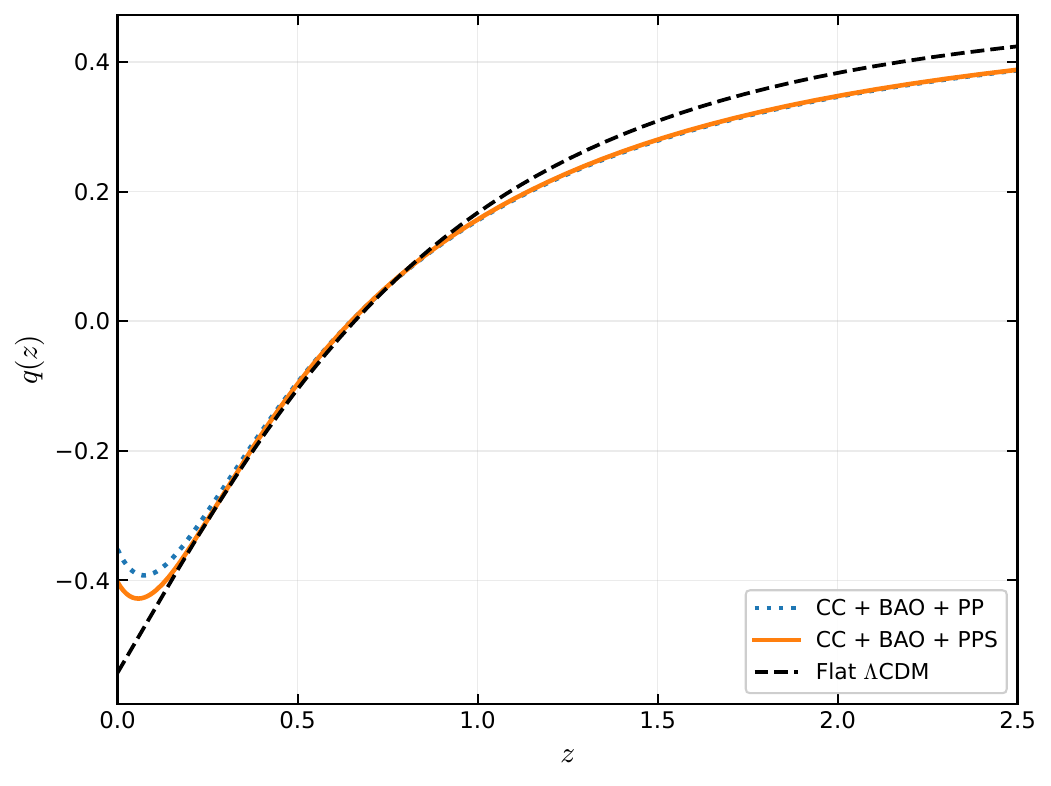}
  \caption{Deceleration parameter $q(z)$ as a function of redshift $z$ for the
$\mathrm{FRS}_{\mathrm{EXP}}$ framework under $\text{CC}+\text{BAO}+\text{PP}$ and $\text{CC}+\text{BAO}+\text{PPS}$ datasets, compared to the flat $\Lambda\text{CDM}$ baseline.}
  \label{fig:q_exp}
\end{minipage}
\label{fig:q_eval}
\end{figure}

The deceleration parameter $q(z)$ tracks the transition from early matter-dominated decelerated expansion ($q > 0$) to late-time cosmic acceleration ($q < 0$), with the transition redshift $z_t$ defined by $q(z_t) = 0$. For a standard flat $\Lambda\text{CDM}$ universe the present-day deceleration evaluates to $q_0 \approx -0.55$, with a characteristic transition at $z_t \approx 0.67$. For the $\mathrm{FRS}_{\mathrm{LOG}}$, the uncalibrated $\text{CC}+\text{BAO}+\text{PP}$ fit yields $q_0 = -0.4208$ and $z_t = 0.6931$, whereas incorporating the SH0ES distance anchor ($\text{CC}+\text{BAO}+\text{PPS}$) slightly deepens current acceleration to $q_0 = -0.4662$ while keeping the transition virtually intact at $z_t = 0.6921$. In comparison, the $\mathrm{FRS}_{\mathrm{EXP}}$ framework predicts a milder present acceleration of $q_0 = -0.3521$ and $q_0 = -0.4027$ across the respective datasets, yet undergoes an earlier transition at $z_t \approx 0.65$ ($z_t = 0.6486$ and $z_t = 0.6490$). As illustrated in figure~\ref {fig:q_log} and ~\ref {fig:q_exp} , both FRS parametrizations approach the matter-dominated kinematic limit, $q \to 0.5$ at high redshift. exponential parametrisation exhibits a low-redshift feature around  $z \lesssim 0.1$, after which \(q(z)\) evolves smoothly towards the higher-redshift decelerating regime.

\subsubsection{Jerk Parameter Departures}
\label{sec:jerk}

\begin{figure}[H]
\centering
\begin{minipage}[t]{0.48\textwidth}
  \centering
  \includegraphics[width=\textwidth]{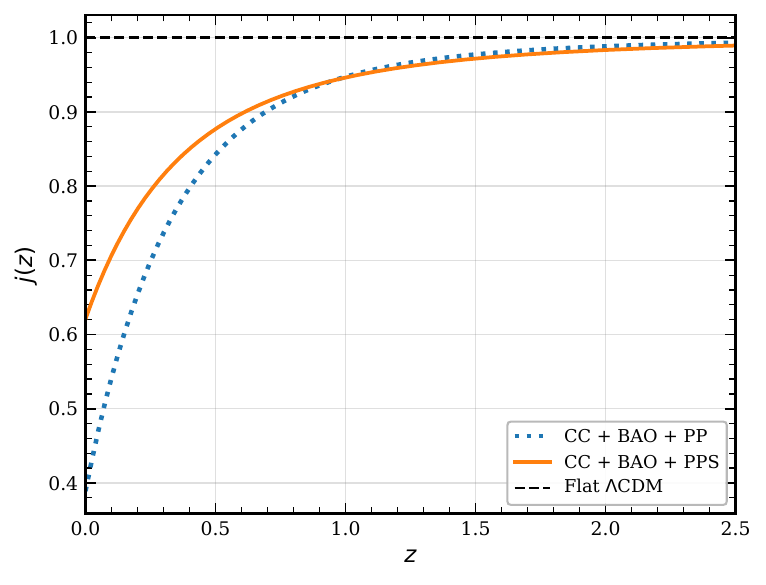}
  \caption{Evolution of the cosmic jerk parameter $j(z)$ for the $\mathrm{FRS}_{\mathrm{LOG}}$  framework. The dashed line denotes the flat $\Lambda\text{CDM}$ benchmark $j = 1$.}
  \label{fig:j_log}
\end{minipage}
\hfill
\begin{minipage}[t]{0.48\textwidth}
  \centering
  \includegraphics[width=\textwidth]{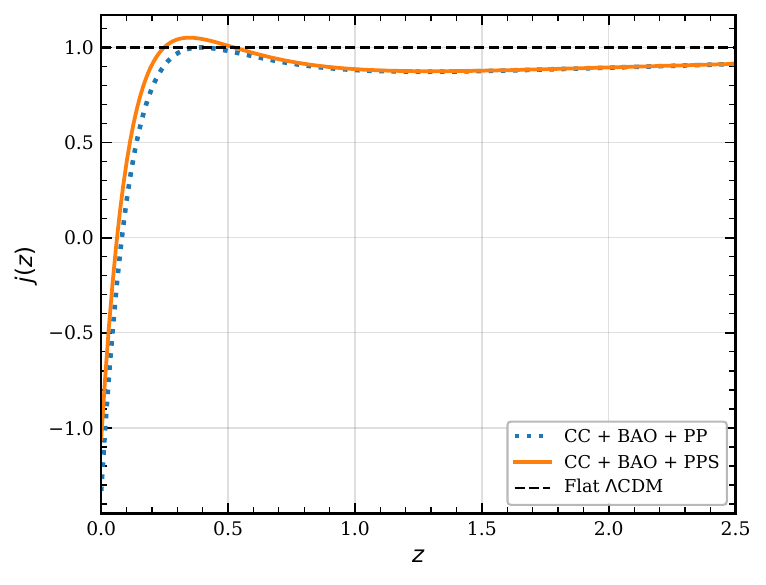}
  \caption{Evolution of the cosmic jerk parameter $j(z)$ for the $\mathrm{FRS}_{\mathrm{EXP}}$ framework, highlighting the late-time dip and low-redshift overshoot.}
  \label{fig:j_exp}
\end{minipage}
\end{figure}

Kinematic departures from $j(z)=1$ provide a useful diagnostic of
deviations from the standard $\Lambda\mathrm{CDM}$ expansion history,
with potential implications for dynamical dark energy or modified
gravity scenarios. As illustrated in Figs.~\ref{fig:j_log} and
\ref{fig:j_exp}, both FRS models exhibit appreciable departures from
the concordance baseline at low redshift. For the $\mathrm{FRS}_{\mathrm{LOG}}$
framework, the present-day jerk is suppressed, yielding
$j_0=0.3891$ for $\mathrm{CC}+\mathrm{BAO}+\mathrm{PP}$ and
$j_0=0.6208$ for $\mathrm{CC}+\mathrm{BAO}+\mathrm{PPS}$. In both
cases, $j(z)$ increases monotonically with redshift and approaches the
canonical $\Lambda\mathrm{CDM}$ value, $j=1$, towards high redshift.
In contrast, the $\mathrm{FRS}_{\mathrm{EXP}}$ model exhibits a pronounced negative
excursion at the present epoch, with $j_0=-1.3318$ and
$j_0=-1.0558$ for the respective dataset combinations. The jerk
parameter then rises sharply, crossing $j=0$ at approximately
$z\sim0.15$, followed by a mild overshoot above the $\Lambda\mathrm{CDM}$
benchmark, reaching $j\approx1.05$ around $z\sim0.3$--$0.4$.
At higher redshifts, the $\mathrm{FRS}_{\mathrm{EXP}}$ model settles into a persistent
sub-unity regime with $j(z)\simeq0.9$, remaining below the
$\Lambda\mathrm{CDM}$ reference over the plotted intermediate and
high-redshift range. These contrasting behaviours provide a clear
kinematic distinction between the two FRS formulations.

\subsubsection{Snap Parameter Evolution}
\label{sec:snap}

\begin{figure}[H]
\centering
\begin{minipage}[t]{0.48\textwidth}
  \centering
  \includegraphics[width=\textwidth]{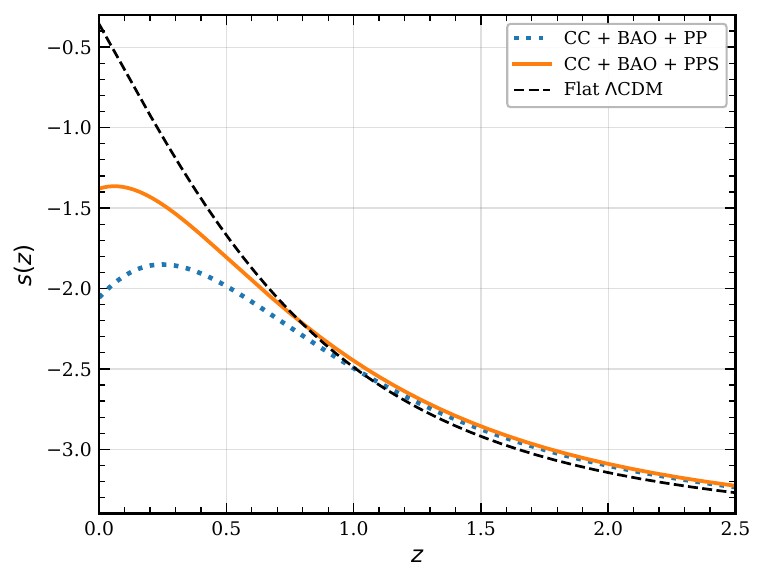}
  \caption{Evolution of the cosmic snap parameter $s(z)$ for the $\mathrm{FRS}_{\mathrm{LOG}}$ framework under both dataset combinations.}
  \label{fig:s_log}
\end{minipage}
\hfill
\begin{minipage}[t]{0.48\textwidth}
  \centering
  \includegraphics[width=\textwidth]{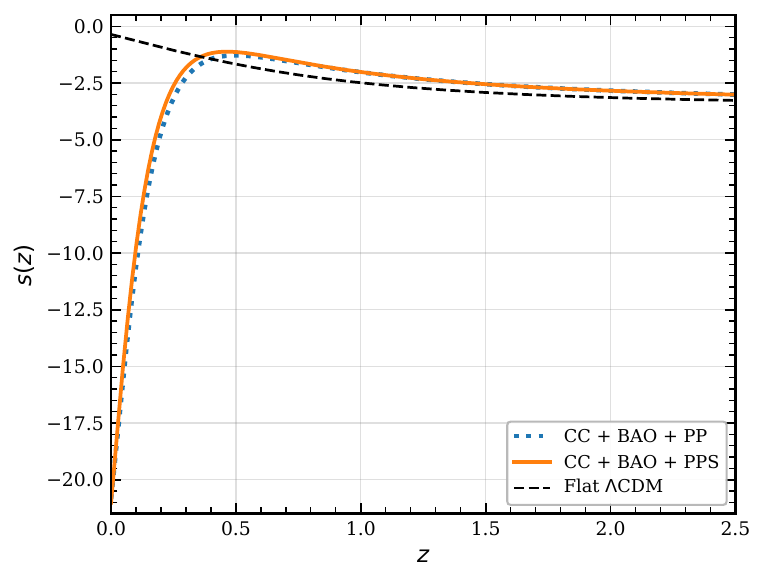}
  \caption{Evolution of the cosmic snap parameter $s(z)$ for the $\mathrm{FRS}_{\mathrm{EXP}}$ framework, illustrating the localized late-time excursion.}
  \label{fig:s_exp}
\end{minipage}
\end{figure}

The snap parameter $s(z)$ involves the fourth time derivative of the scale factor and characterizes higher-order variations in the cosmic expansion history. For a spatially flat $\Lambda\mathrm{CDM}$ model + the present-day value is $s_0 \simeq -0.35$. As shown in figures~\ref{fig:s_log} and~\ref{fig:s_exp}, both FRS formulations exhibit clear departures from this $\Lambda\mathrm{CDM}$ reference, particularly at low redshift. In the $\mathrm{FRS}_{\mathrm{LOG}}$ framework, the snap trajectories remain smooth and negative, with $s_0 = -2.0555$ for $\mathrm{CC}+\mathrm{BAO}+\mathrm{PP}$ and $s_0 = -1.3796$ for $\mathrm{CC}+\mathrm{BAO}+\mathrm{PPS}$. Following their low-redshift behaviour, both trajectories decrease smoothly and approach the $\Lambda\mathrm{CDM}$ prediction towards higher redshifts, with the separation becoming relatively small for $z \gtrsim 1$. In contrast, the 
$\mathrm{FRS}_{\mathrm{EXP}}$ model exhibits a pronounced low-redshift excursion, with $s_0 = -20.8421$ and $s_0 = -21.0060$ for $\mathrm{CC}+\mathrm{BAO}+\mathrm{PP}$ and $\mathrm{CC}+\mathrm{BAO}+\mathrm{PPS}$, respectively. The snap parameter then rises very rapidly from its strongly negative present-day value, reaches a shallow maximum around $z \sim 0.4$, while remaining negative, and subsequently decreases gradually towards the $\Lambda\mathrm{CDM}$ trajectory at higher redshifts. Thus, the Exponential formulation produces a much stronger low-redshift higher-order kinematic excursion than the Logarithmic framework, while both models approach more similar snap behaviour at higher redshift.

\subsubsection{Lerk Parameter Diagnostics}
\label{sec:lerk}

\begin{figure}[H]
\centering
\begin{minipage}[t]{0.48\textwidth}
  \centering
  \includegraphics[width=\textwidth]{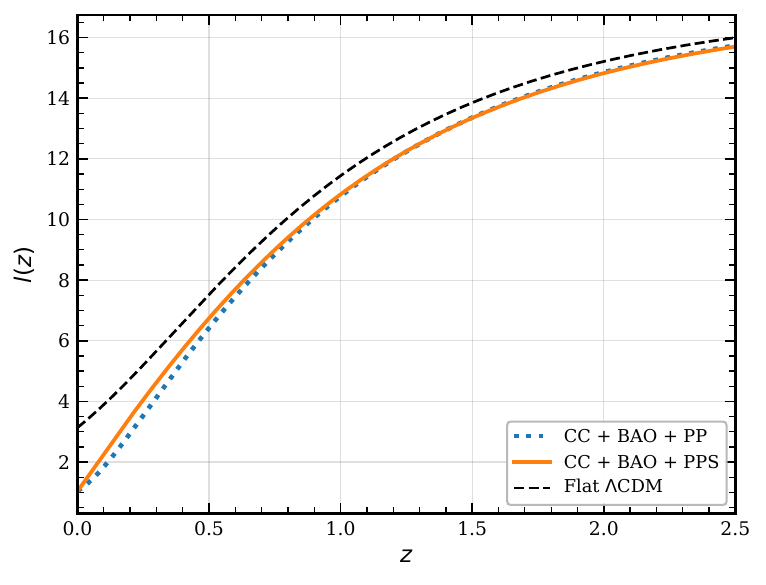}
  \caption{Evolution of the cosmic lerk parameter $l(z)$ for the $\mathrm{FRS}_{\mathrm{LOG}}$  framework as a function of redshift.}
  \label{fig:l_log}
\end{minipage}
\hfill
\begin{minipage}[t]{0.48\textwidth}
  \centering
  \includegraphics[width=\textwidth]{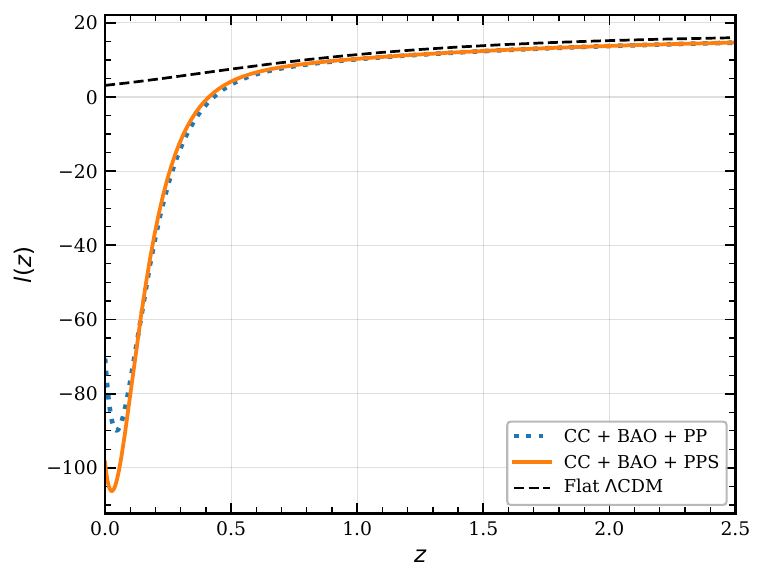}
  \caption{Evolution of the cosmic lerk parameter $l(z)$ for the $\mathrm{FRS}_{\mathrm{EXP}}$  framewor as a function of redshift.}
  \label{fig:l_exp}
\end{minipage}
\end{figure}

The cosmic lerk parameter $l(z)$ involves the fifth time derivative of the scale factor and provides a high-order kinematic diagnostic of the cosmic expansion history. For a spatially flat $\Lambda\mathrm{CDM}$ baseline the present-day reference value is $l_0 \simeq 3.12$. As illustrated in figures~\ref{fig:l_log} and~\ref{fig:l_exp}, the two FRS formulations exhibit contrasting low-redshift behaviour. The $\mathrm{FRS}_{\mathrm{LOG}}$  model remains positive and regular throughout the redshift range considered, with suppressed present-day values of $l_0 = 1.0567$ for $\mathrm{CC}+\mathrm{BAO}+\mathrm{PP}$ and $l_0 = 1.0430$ for $\mathrm{CC}+\mathrm{BAO}+\mathrm{PPS}$. Starting from these lower present-day values, $l(z)$ increases monotonically with redshift and remains slightly below the $\Lambda\mathrm{CDM}$ baseline while gradually approaching its higher-redshift behaviour. In contrast, the $\mathrm{FRS}_{\mathrm{EXP}}$ framework exhibits a pronounced negative excursion at low redshift, with $l_0 = -70.5521$ and $-98.6101$ for $\mathrm{CC}+\mathrm{BAO}+\mathrm{PP}$ and $\mathrm{CC}+\mathrm{BAO}+\mathrm{PPS}$, respectively. As shown in figure~\ref{fig:l_exp}, $l(z)$ develops a sharp localized minimum near the present epoch ($z \lesssim 0.05$), followed by a rapid recovery towards positive values. At higher redshifts, the magnitude of this deviation decreases substantially, and the curves progressively approach the general $\Lambda\mathrm{CDM}$ trend. Thus, the $\mathrm{FRS}_{\mathrm{EXP}}$  formulation exhibits a substantially stronger fifth-order kinematic deviation in the low-redshift regime than the $\mathrm{FRS}_{\mathrm{LOG}}$  formulation, while the difference between the two frameworks becomes less pronounced towards higher redshift.

\begin{table}[H]
\centering
\small
\renewcommand{\arraystretch}{1.25}
\begin{tabular}{llccccc}
\hline\hline
\textbf{Model} & \textbf{Dataset} & $\boldsymbol{q_0}$ & $\boldsymbol{j_0}$ & $\boldsymbol{s_0}$ & $\boldsymbol{l_0}$ & $\boldsymbol{z_t}$ \\
\hline
$\text{FRS}_{\text{LOG}}$ & $\text{CC}+\text{BAO}+\text{PP}$  & $-0.4208$ & $\phantom{-}0.3891$ & $\phantom{0}-2.0555$ & $\phantom{00}1.0567$ & $0.6931$ \\
$\text{FRS}_{\text{LOG}}$ & $\text{CC}+\text{BAO}+\text{PPS}$ & $-0.4662$ & $\phantom{-}0.6208$ & $\phantom{0}-1.3796$ & $\phantom{00}1.0430$ & $0.6921$ \\
$\text{FRS}_{\text{EXP}}$ & $\text{CC}+\text{BAO}+\text{PP}$  & $-0.3521$ & $-1.3318$           & $-20.8421$          & $-70.5521$           & $0.6486$ \\
$\text{FRS}_{\text{EXP}}$ & $\text{CC}+\text{BAO}+\text{PPS}$ & $-0.4027$ & $-1.0558$           & $-21.0060$          & $-98.6101$           & $0.6490$ \\
\hline\hline
\end{tabular}
\caption{Reconstructed present-day cosmographic parameters ($q_0, j_0, s_0, l_0$) and deceleration transition redshift $z_t$ for the FRS gravity models across the joint cosmological datasets, evaluated at the posterior best-fit parameter values.}
\label{tab:cosmography_summary}
\end{table}

\subsection{Geometrical Diagnostics and Effective Dark Energy Evolution}
\label{sec:diagnostics_and_evolution}

Beyond the deceleration history, we map the background geometry of the Finsler--Randers--Sasaki (FRS) models using two complementary kinematic diagnostics: the statefinder pair $\{s_{\rm sf}, r\}$ and the $\{q, r\}$ phase plane. Because these quantities depend on third and higher order derivatives of the scale factor, they probe subtle departures from standard expansion that $q(z)$ alone cannot resolve. Alongside these geometrical diagnostics, we reconstruct the effective dark-energy equation of state $w_{\rm DE}(z)$ and the fractional energy densities directly from the best-fit expansion rate to examine the dynamical behavior of the cosmic fluid.


\subsubsection{Statefinder Diagnostics and Phase-Space Trajectories}
\label{subsec:statefinder_trajectories}

\begin{figure}[H]
\centering
\begin{minipage}[t]{0.48\textwidth}
    \centering
    \includegraphics[width=\textwidth]{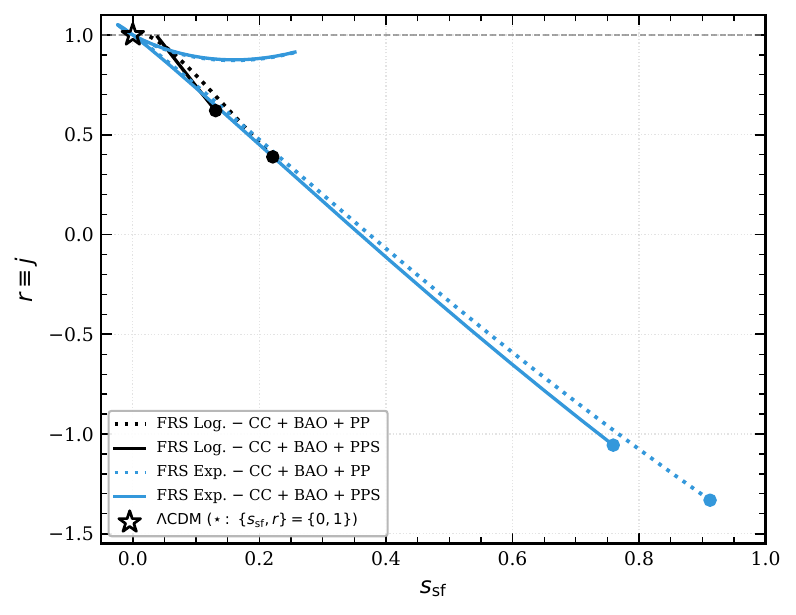}
    \caption{Statefinder trajectories in the $\{s_{\rm sf},r\}$ plane for the $\mathrm{FRS}_{\mathrm{LOG}}$  and $\mathrm{FRS}_{\mathrm{EXP}}$  models. The open star denotes the flat $\Lambda\text{CDM}$ reference point $\{s_{\rm sf},r\}=\{0,1\}$, while circles indicate the present-day values at $z=0$.}
    \label{fig:rs_plane}
\end{minipage}
\hfill
\begin{minipage}[t]{0.48\textwidth}
    \centering
    \includegraphics[width=\textwidth]{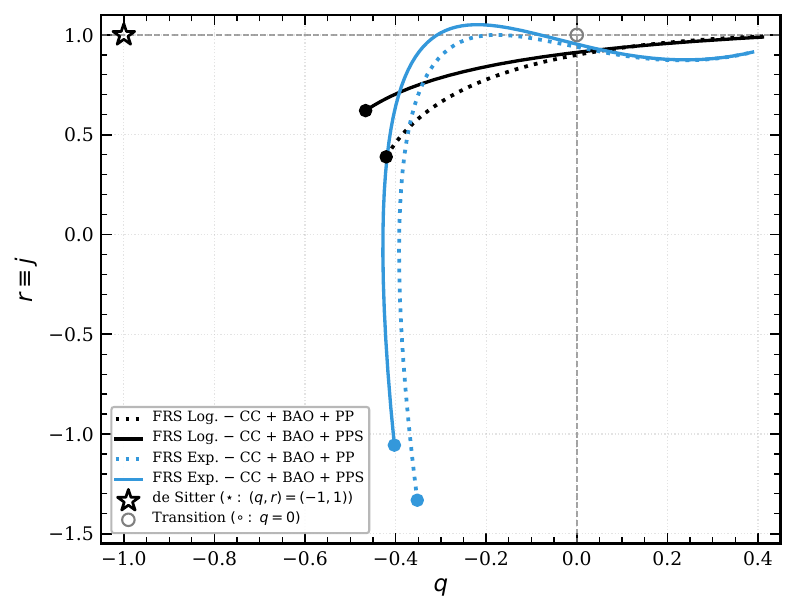}
    \caption{Cosmographic trajectories in the $\{q,r\}$ phase plane for the $\mathrm{FRS}_{\mathrm{LOG}}$  and $\mathrm{FRS}_{\mathrm{EXP}}$  models. The open star denotes the de Sitter reference point $(q,r)=(-1,1)$.}
    \label{fig:rq_plane}
\end{minipage}
\end{figure}

The geometrical statefinder pair $\{s_{\rm sf},r\}$ provides a useful kinematic diagnostic for distinguishing different cosmic expansion histories beyond the deceleration parameter, with the spatially flat $\Lambda\mathrm{CDM}$ model represented by the fixed point $\{s_{\rm sf},r\}_{\Lambda\mathrm{CDM}}=\{0,1\}$. Figure~\ref{fig:rs_plane} illustrates the reconstructed statefinder trajectories obtained from the marginalized central parameter values. For the $\mathrm{FRS}_{\mathrm{LOG}}$  model, the present-day coordinates lie in the region $s_{\rm sf}>0$ and $r<1$, with $(s_{{\rm sf},0},r_0)=(0.2212,0.3891)$ for $\mathrm{CC}+\mathrm{BAO}+\mathrm{PP}$ and $(0.1308,0.6208)$ for $\mathrm{CC}+\mathrm{BAO}+\mathrm{PPS}$. Towards higher redshift, both trajectories evolve smoothly towards the $\Lambda\mathrm{CDM}$ reference point. In contrast, the $\mathrm{FRS}_{\mathrm{EXP}}$ framework exhibits a substantially larger departure in the statefinder plane, with present-day values $r_0=-1.3318$ and $-1.0558$, corresponding to $s_{{\rm sf},0}=0.9121$ and $0.7591$ for $\mathrm{CC}+\mathrm{BAO}+\mathrm{PP}$ and $\mathrm{CC}+\mathrm{BAO}+\mathrm{PPS}$, respectively. The trajectories display a pronounced downward excursion at low redshift before evolving towards the region of the $\Lambda\mathrm{CDM}$ reference point at higher redshift.

The complementary $\{q,r\}$ phase plane is shown in Fig.~\ref{fig:rq_plane}, illustrating the evolution from the decelerating regime ($q>0$) through the transition boundary ($q=0$) into the present accelerated regime ($q<0$). All four FRS cases have negative present-day deceleration parameters, with $q_0=-0.4208$ and $-0.4662$ for the $\mathrm{FRS}_{\mathrm{LOG}}$  model and $q_0=-0.3521$ and $-0.4027$ for the $\mathrm{FRS}_{\mathrm{EXP}}$ model, corresponding respectively to $\mathrm{CC}+\mathrm{BAO}+\mathrm{PP}$ and $\mathrm{CC}+\mathrm{BAO}+\mathrm{PPS}$. The $\mathrm{FRS}_{\mathrm{LOG}}$  trajectories evolve smoothly in the $\{q,r\}$ plane towards the de Sitter reference point $(q,r)=(-1,1)$. By contrast, the $\mathrm{FRS}_{\mathrm{EXP}}$ trajectories exhibit a pronounced low-redshift excursion, with $r$ decreasing to strongly negative values near the present epoch before increasing rapidly and attaining values slightly above unity at intermediate negative $q$. At higher redshifts, the trajectories gradually approach the general $\Lambda\mathrm{CDM}$ behaviour. Over the plotted interval $0\leq z\leq2.5$, all trajectories remain finite and continuous, indicating regular kinematic evolution within the redshift range considered.


\subsubsection{Effective Dark-Energy Behavior and Density Evolution}
\label{subsec:effective_de}

\begin{figure}[H]
\centering
\begin{minipage}[t]{0.48\textwidth}
    \centering
    \includegraphics[width=\textwidth]{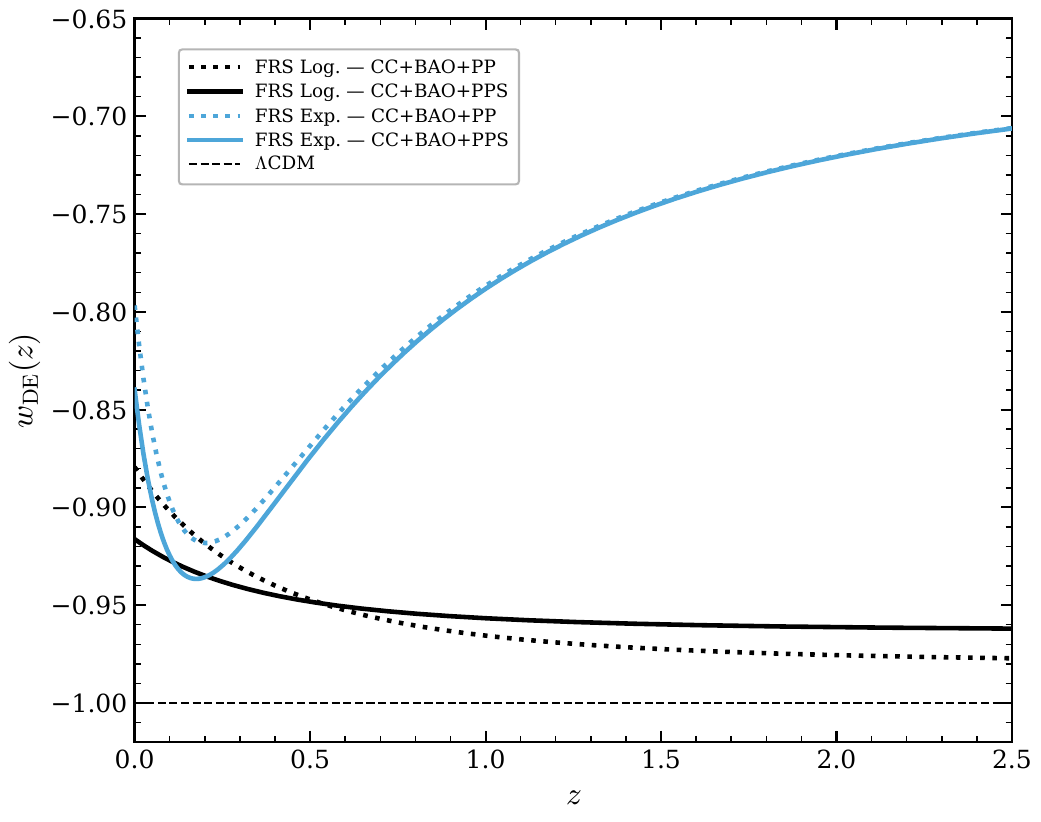}
    \caption{Redshift evolution of the effective dark energy equation of state $w_{\rm DE}(z)$ for the $\mathrm{FRS}_{\mathrm{LOG}}$ and $\mathrm{FRS}_{\mathrm{EXP}}$ models. The horizontal dashed line indicates the cosmological-constant value $w_{\rm DE} = -1$.}
    \label{fig:w_de}
\end{minipage}
\hfill
\begin{minipage}[t]{0.48\textwidth}
    \centering
    \includegraphics[width=\textwidth]{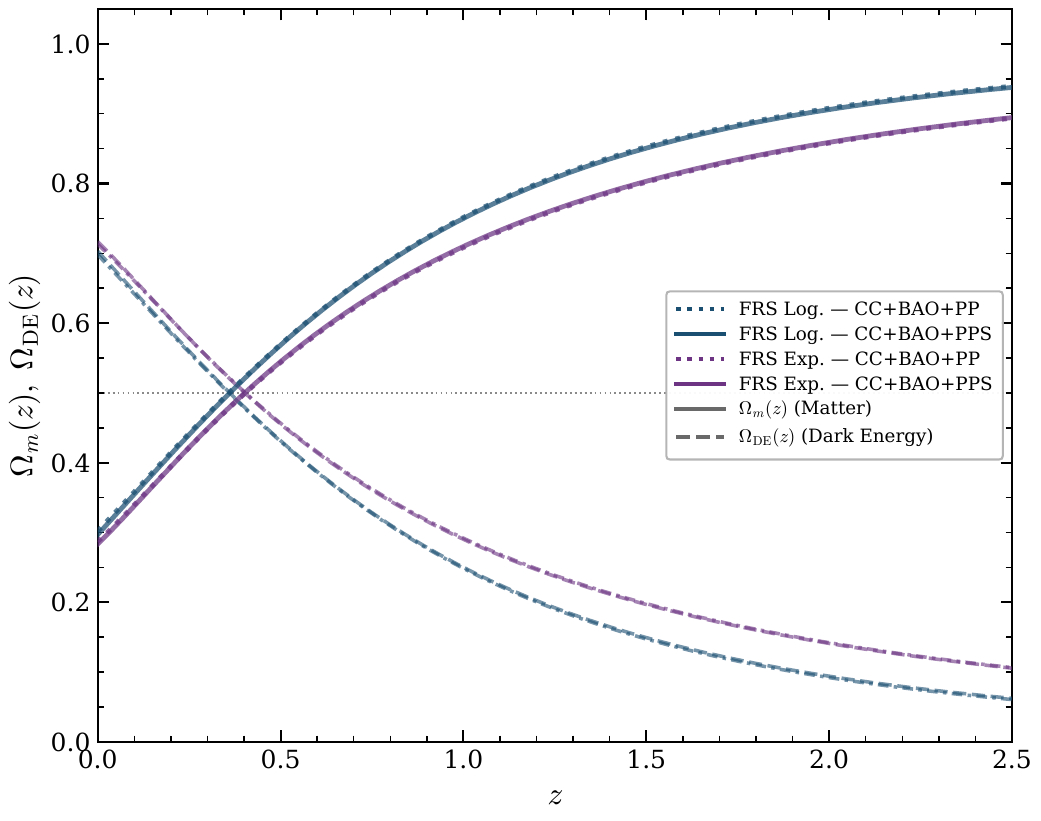}
    \caption{Evolution of the fractional matter density $\Omega_m(z)$ and effective dark energy density $\Omega_{\rm DE}(z)$ for the FRS models. The horizontal dotted line denotes the equal-density condition $\Omega_m = \Omega_{\rm DE} = 0.5$.}
    \label{fig:omega_evolution}
\end{minipage}
\end{figure}

The geometric modifications in the Finsler--Randers--Sasaki framework can be mapped at the background level to an effective dark-energy component whose equation of state $w_{\rm DE}(z)$ is determined by the cosmic expansion rate and its evolution. Evaluating this relation at $z=0$ using the marginalized central parameter values yields quintessence-like values in all cases: $w_{{\rm DE},0}=-0.8794$ and $-0.9163$ for the $\mathrm{FRS}_{\mathrm{LOG}}$  with $\mathrm{CC}+\mathrm{BAO}+\mathrm{PP}$ and $\mathrm{CC}+\mathrm{BAO}+\mathrm{PPS}$, respectively, while the corresponding values for the $\mathrm{FRS}_{\mathrm{EXP}}$  model are $-0.7967$ and $-0.8398$, as summarized in table~\ref{tab:cosmo_dynamics}. As shown in figure~\ref{fig:w_de}, both models remain above the phantom divide, $w_{\rm DE}>-1$, throughout the interval $0\leq z\leq2.5$. Their redshift evolution nevertheless differs significantly: the $\mathrm{FRS}_{\mathrm{LOG}}$ model decreases smoothly from its present-day value and approaches a high-redshift plateau around $w_{\rm DE}\simeq-0.96$ to $-0.98$, whereas the $\mathrm{FRS}_{\mathrm{EXP}}$ model exhibits a localized minimum near $z\sim0.25$ followed by a rapid rise towards less negative values, reaching approximately $w_{\rm DE}\simeq-0.71$ at the upper end of the plotted redshift range.

The corresponding fractional matter and effective dark-energy density evolutions, $\Omega_m(z)$ and $\Omega_{\rm DE}(z)$, are shown in figure~\ref{fig:omega_evolution}. Both FRS frameworks approach the matter-dominated regime at higher redshifts, with $\Omega_m\rightarrow1$ and $\Omega_{\rm DE}\rightarrow0$. The equal-density condition, $\Omega_m=\Omega_{\rm DE}=0.5$, is reached at approximately $z\simeq0.35$--$0.45$ across the considered dataset combinations. Below this redshift, the effective dark-energy component becomes increasingly dominant, reaching $\Omega_{\rm DE,0}\simeq0.70$ at the present epoch and accompanying the transition to accelerated cosmic expansion.


\begin{table}[H]
\centering
\caption{Present-day cosmographic and effective dark-energy parameters reconstructed from the marginalized central values for the FRS Logarithmic and FRS Exponential models.}
\label{tab:cosmo_dynamics}
\vspace{0.2cm}
\begin{tabular}{llcccc}
\hline\hline
Model & Dataset & $q_0$ & $r_0$ & $s_{{\rm sf},0}$ & $w_{{\rm DE},0}$ \\
\hline
$\mathrm{FRS}_{\mathrm{LOG}}$  & $\text{CC}+\text{BAO}+\text{PP}$  & $-0.4208$ & $+0.3891$ & $+0.2212$ & $-0.8794$ \\
$\mathrm{FRS}_{\mathrm{LOG}}$  & $\text{CC}+\text{BAO}+\text{PPS}$ & $-0.4662$ & $+0.6208$ & $+0.1308$ & $-0.9163$ \\
\hline
$\mathrm{FRS}_{\mathrm{EXP}}$  & $\text{CC}+\text{BAO}+\text{PP}$  & $-0.3521$ & $-1.3318$ & $+0.9121$ & $-0.7967$ \\
$\mathrm{FRS}_{\mathrm{EXP}}$  & $\text{CC}+\text{BAO}+\text{PPS}$ & $-0.4027$ & $-1.0558$ & $+0.7591$ & $-0.8398$ \\
\hline\hline
\end{tabular}
\end{table}


\subsection{Statistical Analysis and Model Comparison}
\label{sec:statistical_analysis}

To examine the statistical performance of the Finsler--Randers--Sasaki (FRS) models relative to the standard flat $\Lambda\mathrm{CDM}$ cosmology, we perform a model comparison using the Akaike Information Criterion ($\mathrm{AIC}$)~\cite{Akaike1974} and the Bayesian Information Criterion ($\mathrm{BIC}$)~\cite{Schwarz1978}:
\begin{equation}
\mathrm{AIC}=\chi^2_{\min}+2k,
\qquad
\mathrm{BIC}=\chi^2_{\min}+k\ln(N_{\rm data}),
\label{eq:aic_bic}
\end{equation}
where $\chi^2_{\min}$ denotes the minimum chi-square, $k$ is the number of free parameters and $N_{\rm data}$ is the total number of observational data points. The relative differences are defined as
\begin{equation}
\Delta\mathrm{AIC}
=
\mathrm{AIC}_{\rm model}
-
\mathrm{AIC}_{\Lambda\mathrm{CDM}},
\qquad
\Delta\mathrm{BIC}
=
\mathrm{BIC}_{\rm model}
-
\mathrm{BIC}_{\Lambda\mathrm{CDM}}.
\end{equation}
Following standard conventions~\cite{BurnhamAnderson:2002,KassRaftery:1995}, values of $|\Delta\mathrm{AIC}|\lesssim2$ correspond to only small differences in relative AIC support. For the BIC, we specifically note that $\Delta\mathrm{BIC}>10$ is conventionally associated with very strong evidence against the model having the larger BIC value. The resulting statistical comparison is summarized in table~\ref{tab:statistical_comparison}.

For the uncalibrated $\mathrm{CC}+\mathrm{BAO}+\mathrm{PP}$ compilation, the minimum chi-square values decrease from $\chi^2_{\min}=1414.80$ for $\Lambda\mathrm{CDM}$ to  $1410.33$ for the Logarithmic model and $1411.17$ for the Exponential model and. After accounting for the parameter counts, the corresponding differences are $\Delta\mathrm{AIC}=-0.47$ and $-1.63$, respectively. Both values satisfy $|\Delta\mathrm{AIC}|<2$, indicating only small differences in relative AIC support. The corresponding BIC differences are $\Delta\mathrm{BIC}=10.31$ for the Logarithmic model and $3.76$ for the Exponential model. Thus, only the Logarithmic formulation exceeds the adopted $\Delta\mathrm{BIC}>10$ threshold.

For the $\mathrm{CC}+\mathrm{BAO}+\mathrm{PPS}$ compilation, which incorporates the $\mathrm{SH0ES}$ prior, the minimum chi-square values are $1417.01$  and $1418.09$  for the Logarithmic and Exponential  models, respectively, compared with $1419.44$ for $\Lambda\mathrm{CDM}$. The resulting $\mathrm{AIC}$ differences are $\Delta\mathrm{AIC}=1.57$ and $0.65$, respectively, again remaining within the $|\Delta\mathrm{AIC}|<2$ range. The corresponding $\mathrm{BIC}$ differences are $\Delta\mathrm{BIC}=12.36$ and $6.04$. Hence, the Logarithmic formulation again exceeds the adopted $\Delta\mathrm{BIC}>10$ threshold, whereas the Exponential formulation remains below it. 

Overall, the information-criterion analysis shows that the FRS extensions remain close to flat $\Lambda\mathrm{CDM}$ according to the AIC, with $|\Delta\mathrm{AIC}|<2$ for both observational combinations. The BIC, which imposes a stronger complexity penalty for the additional parameters, gives larger differences. In particular, the FRS Logarithmic formulation exceeds the adopted $\Delta\mathrm{BIC}>10$ threshold for both $\mathrm{CC}+\mathrm{BAO}+\mathrm{PP}$ and $\mathrm{CC}+\mathrm{BAO}+\mathrm{PPS}$, whereas the FRS Exponential formulation remains below this threshold. These results highlight the dependence of the model comparison on the balance between the improvement in the minimum $\chi^2$ and the additional parameter freedom, while the FRS models continue to exhibit distinctive higher-order cosmographic behaviour.

\begin{table}[H]
\centering
\caption{Maximum-likelihood statistics and information-criterion
differences for the flat $\Lambda$CDM, $\mathrm{FRS}_{\mathrm{LOG}}$ and $\mathrm{FRS}_{\mathrm{EXP}}$ 
models. The $\Delta\mathrm{AIC}$ and $\Delta\mathrm{BIC}$ values are
calculated relative to the flat $\Lambda$CDM model for each observational
combination.}
\label{tab:statistical_comparison}

\setlength{\tabcolsep}{5.5pt}
\renewcommand{\arraystretch}{1.25}

\begin{tabular}{llrrrrr}
\toprule
Data combination & Model
& $\chi^2_{\min}$
& AIC
& $\Delta$AIC
& BIC
& $\Delta$BIC \\
\midrule

CC+BAO+PP
& $\Lambda$CDM
& 1414.80
& 1420.80
& 0.00
& 1436.98
& 0.00 \\

& $\mathrm{FRS}_{\mathrm{LOG}}$ 
& 1410.33
& 1420.33
& $-0.47$
& 1447.29
& 10.31 \\

& $\mathrm{FRS}_{\mathrm{EXP}}$ 
& 1411.17
& 1419.17
& $-1.63$
& 1440.74
& 3.76 \\

\midrule

CC+BAO+PPS
& $\Lambda$CDM
& 1419.44
& 1427.44
& 0.00
& 1449.01
& 0.00 \\

& $\mathrm{FRS}_{\mathrm{LOG}}$ 
& 1417.01
& 1429.01
& 1.57
& 1461.37
& 12.36 \\

& $\mathrm{FRS}_{\mathrm{EXP}}$ 
& 1418.09
& 1428.09
& 0.65
& 1455.05
& 6.04 \\

\bottomrule
\end{tabular}
\end{table}

\section{Conclusions}
\label{sec:conclusions}

In this work, we have investigated the late-time cosmological dynamics of
logarithmic and exponential Finsler--Randers--Sasaki (FRS) geometries using
a joint analysis of cosmic chronometers, baryon acoustic oscillations, and
Pantheon+ Type~Ia supernovae. We considered both the uncalibrated
$\mathrm{CC+BAO+PP}$ and SH0ES-calibrated $\mathrm{CC+BAO+PPS}$ data
combinations with flat $\Lambda$CDM used as the
reference cosmology.

The inferred expansion histories show that both FRS parametrizations
provide viable phenomenological descriptions of the late-time background evolution.
For the uncalibrated combination, both models yield values of $H_0$ close
to the $\Lambda$CDM baseline, whereas inclusion of the SH0ES calibration
shifts the inferred Hubble constant to approximately
$H_0\simeq71.6~\mathrm{km\,s^{-1}\,Mpc^{-1}}$ in both FRS models.
The logarithmic model gives a matter density close to
$\Omega_{m0}\simeq0.30$, while the exponential parametrization yields a
somewhat lower matter density. These results indicate that the two FRS
geometries can accommodate the observed late-time expansion while
retaining distinct parameter signatures.

The cosmographic analysis further reveals differences between the two
geometrical constructions. Both models predict accelerated expansion at
the present epoch and a transition from deceleration to acceleration at
$z_t\simeq0.65$--$0.69$. The logarithmic and exponential models exhibit
distinct present-day jerk and statefinder signatures, demonstrating that
higher-order cosmographic diagnostics can distinguish FRS parametrizations
that produce relatively similar constraints on $H_0$. The reconstructed
effective dark-energy sector is quintessence-like at the present epoch,
with $w_{\mathrm{DE},0}>-1$ for both models. The exponential model shows a
larger displacement from the cosmological-constant value
$w_{\mathrm{DE}}=-1$.

We have also assessed the statistical performance of the models using the
Akaike and Bayesian information criteria. Although both FRS models achieve
lower minimum $\chi^2$ values than $\Lambda$CDM for the two data
combinations, the additional parameters are penalized by the information
criteria, particularly by the BIC. The AIC differences are comparatively
small, whereas the positive BIC differences reflect the stronger penalty 
imposed on the additional parameters. Thus, the information criteria provide
a criterion-dependent assessment of the extended FRS parametrizations.

Overall, our results indicate that logarithmic and exponential
Finsler--Randers--Sasaki geometries can provide viable phenomenological
descriptions of the late-time background expansion and exhibit
characteristic signatures in the effective dark-energy and cosmographic
sectors. The differences between the two parametrizations become more
pronounced in higher-order cosmographic quantities than in the inferred
Hubble constant, suggesting that future high-precision observations may
provide useful tests of the underlying FRS geometry.

Future work can extend the present analysis by incorporating
redshift-space distortion and other growth-of-structure observations,
as well as forthcoming high-precision BAO and supernova measurements.
A joint analysis of background expansion and perturbation observables
would provide an important further test of FRS cosmology and its
distinction from the standard $\Lambda$CDM framework.
\acknowledgments
The authors thank the developers of the publicly available computational tools and observational datasets used in this work.




\bibliographystyle{JHEP}

\bibliography{ref}

\end{document}